\documentclass[aps,pra,twocolumn,showpacs,floatfix,longbibliography]{revtex4-1}
\usepackage{graphicx,amsfonts,amssymb,amsmath}
\usepackage[usenames]{color}
\usepackage{textcomp}
\usepackage{esint}
\usepackage[colorlinks,linktocpage,bookmarks=false,citecolor=blue,linkcolor=red,urlcolor=blue]{hyperref}

\newif\ifhyper
\hypertrue
\ifhyper       
\hypersetup{         
  citecolor = {green},
  colorlinks = {true}, 
  urlcolor = {blue} 
}

\allowdisplaybreaks

\begin{document}

\def\rhoeq{\hat\rho_{\rm eq}}

\newcommand{\marge}[1]{\marginpar{\scriptsize #1}}
\newcommand{\remarque}[1]{\marginpar{\scriptsize Remarque}{\it [#1]}}
\newcommand{\new}[1]{{\bf #1}}
\newlength{\textlarg}
\newcommand{\barre}[1]{%
   \settowidth{\textlarg}{#1}
   #1\hspace{-\textlarg}\rule[0.5ex]{\textlarg}{0.5pt}}
\newcommand{\barred}[1]{%
   \settowidth{\textlarg}{#1}
   \red{#1}\hspace{-\textlarg}\rule[0.5ex]{\textlarg}{0.5pt}}
\newcommand{\barblue}[1]{%
   \settowidth{\textlarg}{#1}
   \blue{#1}\hspace{-\textlarg}\rule[0.5ex]{\textlarg}{0.5pt}}

\def\beq{\begin{equation}}
\def\eeq{\end{equation}}
\def\bleq{\begin{eqnarray}}
\def\eleq{\end{eqnarray}} 
\def\bfig{\begin{figure}}
\def\efig{\end{figure}}
\def\bline{\begin{multline}}
\def\eline{\end{multline}}
\def\bremark{\begin{quotation} \noindent \small }
\def\eremark{\end{quotation}}
\def\llbrace{\left\lbrace}
\def\rrbrace{\right\rbrace}
\def\lbraket{\left[}
\def\rbraket{\right]}
\def\llangle{\left\langle}
\def\rrangle{\right\rangle} 

\newcommand{\Tr}{{\rm Tr}} 
\newcommand{\tr}{{\rm tr}} 
\newcommand{\sgn}{{\rm sgn}} 
\newcommand{\mean}[1]{\langle #1 \rangle}
\newcommand{\commu}[2]{[#1,#2]} 
\newcommand{\bra}[1]{\langle#1|}
\newcommand{\ket}[1]{|#1\rangle}
\newcommand{\braket}[2]{\langle #1|#2\rangle}
\newcommand{\dbraket}[3]{\langle #1|#2|#3\rangle}
\newcommand{\tens}[1]{\overleftrightarrow{#1}}  
\newcommand{\vac}{|{\rm vac}\rangle} 
\def\bravac{\langle{\rm vac}|}
\newcommand{\const}{{\rm const}} 
\newcommand{\atanh}{\,{\rm atanh}}
\newcommand{\cotanh}{\,{\rm cotanh}}

\newcommand{\ie}{i.e. }
\newcommand{\iet}{i.e.}
\newcommand{\eg}{e.g. }
\newcommand{\cc}{{\rm c.c.}} 
\newcommand{\hc}{{\rm h.c.}} 
\def\etal{{\it et al. }}

\newcommand{\jhatbf}{\hat {\textbf \j}} 
\newcommand{\Jhatbf}{\hat {\textbf \J}} 
\newcommand{\jhat}{\hat {\jmath}} 
\newcommand{\Jhat}{\hat {J}} 
\newcommand{\jbf}{\textbf j}
\newcommand{\Jbf}{\textbf J}

\def\chibf{\boldsymbol{\chi}}
\def\down{\downarrow}
\def\eps{\epsilon}
\def\gam{\gamma} 
\def\phibf{\boldsymbol{\phi}}
\def\varphibf{\boldsymbol{\varphi}}
\def\varphibfs{\boldsymbol{\varphi}_<}
\def\varphibfl{\boldsymbol{\varphi}_>}
\def\varphis{\varphi_{<}}
\def\varphil{\varphi_{>}}
\def\psibf{\boldsymbol{\psi}}
\def\Ome{\Omega}
\def\omeD{{\omega_D}} 
\def\bfOme{\boldsymbol{\Omega}} 
\def\Omebf{\boldsymbol{\Omega}} 
\def\lamb{\lambda}
\def\Lamb{\Lambda}
\def\sig{\sigma}
\def\Sig{\Sigma}
\def\sigp{{\sigma'}} 
\def\bfsig{\boldsymbol{\sigma}} 
\def\sigbf{\boldsymbol{\sigma}} 
\def\bfSig{\boldsymbol{\Sigma}} 
\def\The{\Theta} 
\def\up{\uparrow}

\def\epsk{\epsilon_{\bf k}} 
\def\xik{\xi_{\bf k}} 
\def\txik{\tilde\xi_{\bf k}} 
\def\xip{\xi_{\bf p}} 
\def\xiq{\xi_{\bf q}} 
\def\xikq{\xi_{{\bf k}+{\bf q}}} 
\def\Ek{E_{\bf k}} 
\def\Ep{E_{\bf p}}
\def\Eq{E_{\bf q}}
\def\Heff{\hat H_{\rm eff}}
\def\Hem{\hat H_{\rm em}}
\def\Hint{\hat H_{\rm int}}
\def\Hloc{\hat H_{\rm loc}}
\def\HMF{\hat H_{\rm MF}}
\def\Sem{S_{\rm em}}
\def\SMF{S_{\rm MF}} 
\def\SHF{S_{\rm HF}} 
\def\SRPA{S_{\rm RPA}} 
\def\Sint{S_{\rm int}} 
\def\Sloc{S_{\rm loc}}
\def\TN{T_{\rm N}} 
\def\TNHF{T^{\rm HF}_{\rm N}} 
\def\Zloc{Z_{\rm loc}} 
\def\ZMF{Z_{\rm MF}} 
\def\ZHF{Z_{\rm HF}} 
\def\ZRPA{Z_{\rm RPA}} 
\def\RPA{{\rm RPA}}
\def\loc{{\rm loc}} 
\def\pp{{\rm pp}}
\def\ph{{\rm ph}} 
\def\ch{{\rm ch}}
\def\sp{{\rm sp}} 
\def\qtf{q_{\rm TF}}
\def\epstf{\eps^{}_{\rm TF}} 
\def\epsrpa{\eps^{}_{\rm RPA}} 
\def\chinnzpp{\chi_{nn}^{0}{}\!\!\!''}

\def\half{\frac{1}{2}}
\def\dhalf{\dfrac{1}{2}}
\def\third{\frac{1}{3}} 
\def\quarter{\frac{1}{4}}

\def\qr{{\bf q}\cdot{\bf r}}
\def\wt{\omega t} 

\def\a{{\bf a}}
\def\b{{\bf b}}
\def\e{{\bf e}}
\def\f{{\bf f}}
\def\g{{\bf g}}
\def\h{{\bf h}}
\def\k{{\bf k}}
\def\l{{\bf l}}
\def\m{{\bf m}}
\def\n{{\bf n}} 
\def\p{{\bf p}} 
\def\q{{\bf q}}
\def\r{{\bf r}}
\def\t{{\bf t}}
\def\u{{\bf u}}
\def\v{{\bf v}}
\def\x{{\bf x}}
\def\y{{\bf y}} 
\def\z{{\bf z}} 
\def\A{{\bf A}}
\def\B{{\bf B}}
\def\D{{\bf D}} 
\def\E{{\bf E}} 
\def\F{{\bf F}} 
\def\H{{\bf H}}  
\def\J{{\bf J}}
\def\K{{\bf K}} 

\def\G{{\bf G}}
\def\L{{\bf L}}
\def\M{{\bf M}}  
\def\O{{\bf O}} 
\def\P{{\bf P}} 
\def\Q{{\bf Q}} 
\def\R{{\bf R}}
\def\S{{\bf S}}
\def\epsbf{\boldsymbol{\epsilon}}
\def\mubf{\boldsymbol{\mu}}
\def\nablabf{\boldsymbol{\nabla}}
\def\rhobf{\boldsymbol{\rho}}
\def\sigmabf{\boldsymbol{\sigma}} 
\def\Pibf{\boldsymbol{\Pi}}
\def\pibf{\boldsymbol{\pi}}

\def\para{\parallel}
\def\kpara{{k_\parallel}}
\def\kperp{{k_\perp}} 
\def\kperpp{{k_\perp'}} 
\def\qperp{{q_\perp}} 
\def\tperp{{t_\perp}} 

\def\w{\omega}
\def\wn{\omega_n}
\def\wm{\omega_m}
\def\wnu{\omega_\nu}
\def\wp{\omega_p} 
\def\dmu{{\partial_\mu}}
\def\dnu{{\partial_\nu}}
\def\dl{{\partial_l}}  
\def\dt{\partial_t} 
\def\tdt{\tilde\partial_t}
\def\dk{\partial_k}
\def\tdk{\tilde\partial_k}
\def\dx{\partial_x}
\def\dy{\partial_y} 
\def\dtau{{\partial_\tau}}  
\def\det{{\rm det}} 
\def\Pf{{\rm Pf}}

\def\dsum{\displaystyle \sum}
\def\dint{\displaystyle \int} 
\def\intt{\int_{-\infty}^\infty dt} 
\def\inttp{\int_{-\infty}^\infty dt'} 
\def\intk{\int_{\bf k}} 
\def\intkd{\int \frac{d^dk}{(2\pi)^d}}
\def\intq{\int_{\bf q}} 
\def\intr{\int d^dr}  
\def\dintr{\displaystyle \int d^dr} 
\def\intrp{\int d^dr'}
\def\dinttau{\displaystyle \int_0^\beta d\tau}
\def\dinttaup{\displaystyle \int_0^\beta d\tau'}
\def\inttau{\int_0^\beta d\tau}
\def\inttaup{\int_0^\beta d\tau'}
\def\intx{\int d^{d+1}x} 
\def\inttaur{\int_0^\beta d\tau \int d^dr}
\def\intinf{\int_{-\infty}^\infty}
\def\dinttaur{\displaystyle \int_0^\beta d\tau \int d^dr}
\def\dintinf{\displaystyle \int_{-\infty}^\infty}
\def\intw{\int_{-\infty}^\infty \frac{d\w}{2\pi}}
\def\sumr{\sum_{\bf r}} 

\def\calA{{\cal A}}
\def\calB{{\cal B}} 
\def\calC{{\cal C}} 
\def\dt{\partial_t}
\def\calD{{\cal D}}
\def\calF{{\cal F}} 
\def\calG{{\cal G}}
\def\calH{{\cal H}}
\def\calI{{\cal I}}
\def\calJ{{\cal J}}
\def\calK{{\cal K}}
\def\calL{{\cal L}} 
\def\calM{{\cal M}} 
\def\calN{{\cal N}}
\def\calO{{\cal O}}
\def\calP{{\cal P}}  
\def\calR{{\cal R}} 
\def\calS{{\cal S}}
\def\calT{{\cal T}}
\def\calU{{\cal U}}
\def\calX{{\cal X}} 
\def\calY{{\cal Y}} 
\def\calZ{{\cal Z}} 

\def\calbfB{{\bf \cal B}}
\def\calbfF{{\bf \cal F}}

\def\tT{{\tilde T}}
\def\talpha{{\tilde\alpha}}
\def\tdelta{{\tilde\delta}}
\def\teta{{\tilde\eta}} 
\def\tlamb{{\tilde\lambda}}
\def\tmu{{\tilde\mu}}
\def\tphibf{{\tilde\phibf}}
\def\trho{{\tilde\rho}}
\def\tvarphibf{{\tilde\varphibf}} 
\def\tw{{\tilde\omega}}
\def\twn{{\tilde\omega_n}}

\def\asinh{{\rm asinh}} 
\graphicspath{{./figuresv4/}}

\def\TcMF{T_c^{\rm MF}} 
\def\qmax{q_{\rm max}} 

\title{First-order phase transitions in spinor Bose gases and frustrated magnets}

\author{T. Debelhoir}
\author{N. Dupuis}
\affiliation{Laboratoire de Physique Th\'eorique de la Mati\`ere Condens\'ee, UPMC, 
CNRS UMR 7600, Sorbonne Universit\'es, 4 Place Jussieu, 
75252 Paris Cedex 05, France}

\date{November 14, 2016} 

\begin{abstract}
We show that phase transitions in spin-one Bose gases and stacked triangular Heisenberg antiferromagnets -- an example of frustrated magnets with competing interactions -- are described by the same Landau-Ginzburg-Wilson Hamiltonian with O(3)$\times$O(2) symmetry. In agreement with previous nonperturbative-renormalization-group studies of the three-dimensional O(3)$\times$O(2) model, we find that the transition from the normal phase to the superfluid ferromagnetic phase in a spin-one Bose gas is weakly first order and shows pseudoscaling behavior. The (nonuniversal) pseudoscaling exponent $\nu$ is fully determined by the scattering lengths $a_0$ and $a_2$. We provide estimates of $\nu$ in $^{87}$Rb, $^{41}$K and $^7$Li atom gases which can be tested experimentally. We argue that pseudoscaling comes from either a crossover phenomena due to proximity of the O(6) Wilson-Fisher fixed point ($^{87}$Rb and $^{41}$K) or the  existence of two unphysical fixed points (with complex coordinates) which slow down the RG flow ($^7$Li). These unphysical fixed points are a remnant of the chiral and antichiral fixed points that exist in the O($N$)$\times$O(2) model when $N$ is larger than $N_c\simeq 5.3$ (the transition being then second order and controlled by the chiral fixed point). Finally, we discuss a O(2)$\times$O(2) lattice model and show that our results, even though we find the transition to be first order, are compatible with Monte Carlo simulations yielding an apparent second-order transition.  
\end{abstract}
\maketitle

\section{Introduction}

Ultracold dilute atomic gases are ideal laboratories for the realization of (quantum) simulators thus providing an alternate approach to numerical simulations for understanding minimal models of condensed-matter systems~\cite{Bloch08}. This is due to the perfect control and tunability of the interactions in these systems. In a recent paper~\cite{Debelhoir16a}, we have shown that phase transitions in three-dimensional stacked triangular Heisenberg  antiferromagnets (STHAs) --an example of frustrated magnets with competing interactions-- can be simulated by spinor Bose gases, which opens up the possibility to solve the long-standing controversy about the nature (second or weakly first order) of phase transitions in these frustrated magnets. 

Both STHAs and spin-one bosons are described by a Landau-Ginzburg-Wilson with O(3)$\times$O(2) symmetry. Whereas perturbative renormalization-group (RG) calculations in fixed dimension $d=3$ predict a second-order phase transition~\cite{Pelissetto01a,Calabrese02,Calabrese03,Calabrese04a}, perturbative RG near $d=4$~\cite{Garel76,Bailin77,Yosefin85,Antonenko95a,Holovatch04,Calabrese04} and the nonperturbative renormalization group (NPRG)~\cite{Tissier00,Tissier03,Delamotte04,Delamotte16} find a first-order phase transition. In the latter case however, even though there is no stable fixed point, the RG flow is very slow in a whole region of the coupling constant space due to two unphysical fixed points with complex coordinates~\cite{Zumbach93,Delamotte04}. This implies the possibility to observe pseudoscaling with effective (nonuniversal) exponents on a large temperature range. In spin-one Bose gases, the pseudocritical exponents depend only on the (known) $s$-wave scattering lengths $a_0$ and $a_2$ which, in contrast to STHA materials, allows us to make theoretical predictions that can be tested experimentally~\cite{Debelhoir16a}. 

In this paper we further study the superfluid transition in spin-one boson systems within the NPRG framework. On the one hand, we improve the approach of Ref.~\cite{Debelhoir16a} by taking into account quantum fluctuations. To this end we introduce a two-step NPRG approach: fluctuations with momenta larger than the inverse of the thermal de Broglie wavelength $\lamb_{\rm dB}$ are first integrated out in a simple approximation of the exact NPRG flow equation. This yields a classical O(3)$\times$O(2) model describing thermal fluctuations which is studied within the well-known LPA$'$ approximation (an improvement of the local potential approximation (LPA)). The inclusion of quantum fluctuations removes the dependence of the pseudocritical exponents on the upper momentum cutoff $\Lamb_T\sim\lamb_{\rm dB}^{-1}$ of the classical O(3)$\times$O(2) model, which was the main source of uncertainty in our previous work~\cite{Debelhoir16a}. On the other hand, we study the O($N$)$\times$O(2) model as a function of $N$. For $N\geq N_c\simeq 5.3$, the transition is second order and controlled by the ``chiral'' fixed point which, in the RG flow diagram, coexists with the ``antichiral'', Gaussian and O($2N$) Wilson-Fisher fixed points. When $N=N_c$, the chiral and antichiral fixed points merge; for $N<N_c$ they are replaced by two unphysical fixed points (with complex coordinates) which slow down the RG flow and may induce pseudocritical behavior depending on the values of the scattering lengths $a_0$ and $a_2$. Finally we study a O(2)$\times$O(2) lattice model and compare our results with those of Ref.~\cite{Calabrese04a} where, on the basis of Monte Carlo simulations, it was argued that the transition is second order in some parameter range. 

The paper is organized as follows. In Sec.~\ref{sec_O3O2model} we show that the superfluid transition in spin-one boson systems is described by the O(3)$\times$O(2) model, as in STHAs~\cite{[{A similar analogy between
spin-$\half$ Bose systems and stacked frustrated XY antiferromagnets has been pointed out by }]Ceccarelli15}. The two-step NPRG approach is described in Sec.~\ref{sec_nprg}. By computing the Gibbs free energy we show that the transition from the normal phase to the superfluid ferromagnetic phase is weakly first order. The correlation length increases with a pseudocritical exponent $\nu$ on a large temperature range before saturating at the transition temperature. $\nu$ is computed for $^{87}$Rb, $^{41}$K and $^7$Li atom gases. For $^{87}$Rb and $^{41}$K, pseudoscaling can be explained by a crossover phenomenon due to the proximity of the O(6) Wilson-Fisher fixed point. In the case of $^7$Li, pseudoscaling is due to the presence of two unphysical fixed points. The O($N$)$\times$O(2) model with $3\leq N\leq 6$ is discussed in Sec.~\ref{sec_ONO2} and the O(2)$\times$O(2) lattice model in Sec.~\ref{sec_O2lat}. The experimental measurement of the pseudocritical exponent $\nu$ in spin-one Bose gases is discussed in the conclusion.

\section{O(3)$\times$O(2) model} 
\label{sec_O3O2model}

In this section we show that the superfluid transition in spin-one boson systems is described by
the O(3)$\times$O(2) model. 

Let us consider the Hamiltonian of spin $f=1$ bosons~\cite{Ohmi98,Ho98,Kawaguchi12,*Stamper-Kurn13}. The kinetic energy part is simply
\begin{equation}
\hat H_0 = \int d^3r \sum_{m=-f}^f \left( \frac{1}{2M} \nablabf\hat\psi^\dagger_m \cdot \nablabf\hat\psi_m - \mu \hat\psi^\dagger_m \hat\psi_m \right) , 
\label{H0}
\end{equation}
where $\mu$ is the chemical potential, $M$ the boson mass, and $\hat\psi_m(\r)$ an annihilation operator of a boson at point $\r$ in the spin state $\ket{f,m}$ (we set $\hbar=k_B=1$). The quantum number $m\in[-f,f]$ refers to the spin projection on the $z$ axis. Since the total spin is conserved in a binary collision, the interaction Hamiltonian is determined by three potentials $v^{(F)}(\r,\r')$ where $F=0,1,\cdots,2f$ is the total spin of the colliding particles~\cite{Kawaguchi12}, 
\begin{align}
\Hint ={}& \half \int d^3r \, d^3r' \sum_{F=0}^{2f} v^{(F)}(\r,\r') \nonumber \\ & 
\times \sum_{\calM=-F}^F \hat A^\dagger_{F\calM}(\r,\r') \hat A_{F\calM}(\r,\r') .
\label{Hint}
\end{align} 
The operator 
\begin{equation}
\hat A_{F\calM}(\r,\r') = \sum_{m,m'=-f}^f \braket{F,\calM}{f,m;f,m'} \hat\psi_m(\r) \hat\psi_{m'}(\r') 
\label{AFMdef} 
\end{equation}
annihilates a pair of bosons in the spin state $\ket{F,\calM}$ located at $\r$ and $\r'$. 

A classical Hamiltonian describing the low-energy behavior can be obtained by integrating out fluctuations with momenta larger than the inverse of the thermal de Broglie wavelength $\lamb_{\rm dB}=(2\pi/MT)^{1/2}$. 
Suppose that, starting from Hamiltonian~(\ref{H0},\ref{Hint}),
we integrate out fluctuations down to the momentum scale
$\Lamb_T\sim\sqrt{2MT}\sim\lamb^{-1}_{\rm dB}$. Since $\Lamb_T\gg \Lamb_\mu\sim\sqrt{2M\mu}$ near the 
transition~\footnote{This follows from $Ma_F^2\mu,Ma_F^2T\ll 1$ at low temperature and density, and $M(a_0+5a_2)^2\mu\sim[M(a_0+5a_2)^2T]^{3/2}$ for $T\sim T_c$ [see Eq.~(\ref{muHF})].}, the RG flow at momentum scales larger than $\Lamb_T$ is
effectively in vacuum ($T=0$ and $\mu=0$). In three dimensions, the flow in vacuum 
is controlled by a Gaussian fixed
point and the interactions are irrelevant (the upper critical dimension for the vacuum-superfluid
transition is $d^+_c=2$)~\cite{Sachdev_book}. For momentum 
scales smaller than $1/a_F$, the two-body interaction becomes nearly momentum
independent and equal to $4\pi a_F/m$ where $a_F$ is the $s$-wave scattering length in the total spin channel $F$. Higher-order 
(irrelevant) interactions can be ignored (they yield contributions that are subleading in the small parameter $n a_F^3$; $n$ denotes the density)~\footnote{The two-body interaction is dangerously irrelevant (in the RG sense) and cannot be ignored.}. At momentum scales
smaller than $\Lamb_T$, the contribution of nonclassical modes to the flow is negligible.  
Thus, the renormalized action at the
thermal scale $\Lamb_T\sim\lamb^{-1}_{\rm dB}$ corresponds to a classical field theory
with interaction constants $g_F=4\pi a_F/m$~\footnote{For $|\p|\ll\lamb^{-1}_{\rm dB}$ the number of bosons with momentum $\p$, given by the Bose-Einstein distribution function, is much larger than unity, hence the classical behavior. For an alternative derivation of the classical field theory in spin-zero boson gases, see, e.g., Ref.~\cite{Blaizot08}.}. The Clebsch-Gordan coefficient $\braket{F=1,\calM}{f,m;f,m'}$ being odd in the exchange $m\leftrightarrow m'$, $s$-wave scattering is not possible in the total spin channel $F=1$, so that only $g_0$ and $g_2$ have to be considered. 
In Sec.~\ref{subsec_nprg} we show how the integration of fluctuations with momenta larger than $\Lamb_T$ can be carried out within the NPRG approach. 
  
Thus the phase transition of a spin-one boson gas can be studied from the classical Hamiltonian
\begin{multline}
H = \beta \int d^3r \Biggl\lbrace \sum_{m=-f}^f \left[ \frac{|\nablabf\psi_m|^2}{2M}
 - \mu' |\psi_m|^2 \right] \\
+ \half \sum_{F=0,2} g_F \sum_{\calM=-F}^F A^*_{F\calM}(\r) A_{F\calM}(\r) \Biggr\rbrace ,
\label{ham3}
\end{multline}
where $A_{F\calM}(\r)\equiv A_{F\calM}(\r,\r)$, $\beta=1/T$ and $\psi_m(\r)$ is now a complex field. $\mu'$ denotes a renormalized chemical potential whose value does not matter for our purpose. 
For spin-zero bosons, the classical Hamiltonian has been used to compute the shift in the Bose-Einstein-condensation (BEC) temperature due to interactions~\cite{[{See, e.g., }]Baym01}.

Instead of the basis $\lbrace\ket{1,m}\rbrace$ it is convenient to use the Cartesian basis, defined by $\hat F^\alpha\ket{1,\alpha}=0$ ($\alpha=x,y,z$), where the field $\psibf=(\psi_x,\psi_y,\psi_z)^T$ transforms as a vector under spin rotation ($\hat F^\alpha$ is a spin-one matrix). Using
\begin{equation}
\begin{split} 
|A_{00}|^2 &= \third |\psibf\cdot\psibf|^2 , \\ 
\sum_{\calM=-2}^2 |A_{2\calM}|^2 &= (\psibf^*\cdot\psibf)^2 - \third |\psibf\cdot\psibf|^2 , 
\end{split}
\end{equation}
one obtains the Hamiltonian   
\begin{align}
H ={}& \beta \int d^3r\biggl\lbrace \frac{1}{2M} |\nablabf\psibf|^2 - \mu' |\psibf|^2 \nonumber \\ &
+ \frac{g_2}{2} (\psibf^*\cdot\psibf)^2 
+ \frac{g_0-g_2}{6} |\psibf\cdot\psibf|^2  \biggr\rbrace 
\label{ham4}
\end{align}
which is manifestly invariant under spin inversion and rotation, U(1) (gauge) transformation, and time reversal (complex conjugation) $\Theta$, i.e., under the symmetry group $G={\rm O}(3)\times U(1)\times\Theta$~\cite{Kawaguchi11}. 

If one writes the complex field 
\begin{equation}
\psibf=\sqrt{M/\beta} (\varphibf_1+i \varphibf_2)
\end{equation}
in terms of two real fields $\varphibf_1$ and $\varphibf_2$, one obtains the standard Hamiltonian of the O(3)$\times$O(2) model,
\begin{equation}
H = \int d^3 r \llbrace \half \bigl[(\nablabf \varphibf_1)^2 +(\nablabf \varphibf_2)^2 \bigr] + r \rho + \frac{\lamb_1}{2} \rho^2 + \frac{\lamb_2}{2} \tau \rrbrace 
\label{ham2}
\end{equation}
with 
\begin{equation}
\begin{gathered}
\rho=\half(\varphibf_1^2+\varphibf_2^2), \\
\tau=\quarter( \varphibf_1^2-\varphibf_2^2)^2+(\varphibf_1\cdot\varphibf_2)^2 ,
\end{gathered} 
\label{rtdef} 
\end{equation}
and a momentum cutoff $\Lamb_T$ of the order of the inverse of the thermal de Broglie wavelength $\lamb_{\rm dB}$ (see Sec.~\ref{sec_nprg} and Appendix~\ref{app_rgeq_quantum} for a further discussion), where 
\begin{equation} 
\begin{gathered}
r\equiv-2M\mu' , \\ 
\lamb_1\equiv(4M^2/\beta)g_2 , \quad \lamb_2\equiv(4M^2/3\beta)(g_0-g_2) .
\end{gathered}
\label{hamrel}
\end{equation} 
The symmetry group $G$ is now O(3)$\times$O(2), where O(3) corresponds to a global rotation of $\varphibf_1$ and $\varphibf_2$ whereas the O(2) transformation mixes $\varphibf_1$ and $\varphibf_2$: 
\begin{equation}
\begin{split} 
\varphibf_1' &= \cos \alpha\varphibf_1 - \sin\alpha \varphibf_2 , \\ 
\varphibf_2' &= \pm(\sin \alpha\varphibf_1 + \cos\alpha \varphibf_2).
\end{split}
\label{tO2} 
\end{equation} 
The SO(2) rotation and the $\pm$ sign in~(\ref{tO2}) correspond, respectively, to U(1) rotation and time reversal in the original bosonic picture.  

For $\lamb_2>0$ (the case corresponding to noncollinear spin ordering in the STHA~\cite{note19}), i.e., $g_0>g_2$, the superfluid phase is the so-called ferromagnetic phase~\cite{Kawaguchi12,*Stamper-Kurn13}. For $\lamb_2=0$ the Hamiltonian possesses an O(6) symmetry; the transition is second order and controlled by the O(6) Wilson-Fisher fixed point.

\section{Superfluid transition of spin-one bosons} 
\label{sec_nprg}

We now discuss the NPRG approach to study the superfluid transition in spin-one boson systems. Since at low energy the interactions are fully parameterized by the scattering lengths $a_0$ and $a_2$, we can consider the quantum Hamiltonian 
\begin{align}
\hat H =& \int d^3 r \biggl\{ \frac{1}{2M} \nablabf\hat\psibf{}^\dagger \cdot \nablabf\hat\psibf - \mu \hat\psibf{}^\dagger \hat\psibf \nonumber \\ 
& + \frac{g_{2,\Lamb}}{2} (\hat\psibf{}^\dagger \hat\psibf)^2 + \frac{g_{0,\Lamb}-g_{2,\Lamb}}{6} |\hat\psibf \cdot \hat\psibf|^2 \biggr\} ,
\label{ham5}
\end{align}
where $\hat\psibf=(\hat\psi_x,\hat\psi_y,\hat\psi_z)^T$ is the bosonic operator defined in the Cartesian basis (Sec.~\ref{sec_O3O2model}). The interaction is assumed to be local in space and an upper momentum cutoff $\Lamb$ is implied. In this model, the scattering length $a_F$ is a function of $g_{F,\Lamb}$ and $\Lamb$ (and $M$), and to leading order in $Ma_F^2T$ and $Ma_F^2\mu$ physical quantities can be expressed in terms of $a_F$ with no explicit reference to $g_{F,\Lamb}$ and $\Lamb$. 

The RG procedure is divided into two steps. In the first one, we integrate out fluctuations with momenta larger than $\Lambda_T\sim \lamb^{-1}_{\rm dB}$ within a simple approximation where only a small number of interaction constants are considered. In the second one, we integrate out classical (thermal) fluctuations with momenta smaller than $\Lambda_T$ in a more refined approximation where the full field dependence of the effective potential (or Gibbs free energy) is computed. We will see that the final results are essentially independent of the precise value of the thermal cutoff $\Lamb_T$.

\subsection{NPRG approach} 
\label{subsec_nprg} 

The strategy of the NPRG approach is to build a family of theories indexed by a momentum scale $k$ such that fluctuations are smoothly taken into account as $k$ is lowered from the microscopic scale $\Lamb$ down to 0~\cite{Berges02,Delamotte12,Kopietz_book}. 

Let us consider the action 
\begin{equation}
S = \inttau \llbrace \int d^3r\, \psibf^\dagger(\r,\tau) \dtau \psibf(\r,\tau) + H[\psibf^\dagger,\psibf] \rrbrace ,
\end{equation}
where $\psibf(\r,\tau)$ is a complex vector field and $\tau\in[0,\beta]$ an imaginary time. $H[\psibf^\dagger,\psibf]$ is obtained from Hamiltonian~(\ref{ham5}) by replacing the bosonic operator $\hat\psi{}_\alpha^{(\dagger)}(\r)$ by the complex field $\psi_\alpha^{(*)}(\r,\tau)$ ($\alpha=x,y,z$). To implement the NPRG approach, we add to this action the infrared regulator (for NPRG studies of spin-zero bosons, see Refs.~\cite{Wetterich08,Dupuis07,Dupuis09a,Dupuis09b,Sinner09,Sinner10}) 
\begin{equation}
\Delta S_k[\psibf^\dagger,\psibf] =  \sum_{\p,\wn} \psibf^\dagger(\p,i\wn) R_k(\p) \psibf(\p,i\wn) , 
\end{equation} 
where $\psibf^{(\dagger)}(\p,i\wn)$ is the Fourier transform field of $\psibf^{(\dagger)}(\r,\tau)$ and $\wn=2n\pi T$ ($n$ integer) a bosonic Matsubara frequency. We choose a cutoff function $R_k(\p)$ which does not depend on frequency. 
The partition function
\begin{equation}
Z_k[\J^\dagger,\J] = \int \calD[\psibf^\dagger,\psibf]\, e^{ - S - \Delta S_k + \inttau \int d^3r (\J^\dagger\psibf+\cc) }
\end{equation}
is now $k$ dependent. The scale-dependent effective action
\begin{align}
\Gamma_k[\phibf^\dagger,\phibf] ={}& - \ln Z_k [\J^\dagger,\J]  
+ \inttau \int d^3r\, (\J^\dagger \phibf + \cc) \nonumber \\ &  - \Delta S_k[\phibf^\dagger,\phibf]
\label{Gamdef} 
\end{align}
is defined as a modified Legendre transform of $−\ln Z_k[\J^\dagger,\J]$ which includes the subtraction of $\Delta S_k[\phibf^\dagger,\phibf]$. Here $\phibf^{(\dagger)}(\r,\tau)=\mean{\psibf^{(\dagger)}(\r,\tau)}$ is the order parameter (in the presence of the external complex source $\J$). 

The initial condition of the flow is specified by the microscopic scale $k=\Lamb$ where fluctuations are assumed to be frozen by the $\Delta S_k$ term, so that $\Gamma_\Lambda[\phibf^\dagger,\phibf]=S[\phibf^\dagger,\phibf]$ reproduces mean-field (Bogoliubov) theory. The effective action of the original model is given by $\Gamma_{k=0}$ provided that $R_{k=0}$ vanishes. For a generic value of $k$, the regulator $R_k(\p)$ suppresses fluctuations with momentum $|\p|\lesssim k$ but leaves unaffected those with $|\p|\gtrsim k$. For the most part, we use the theta regulator~\cite{Litim01} 
\begin{equation}
R_k(\p) = \frac{Z_k}{2M} (k^2-\p^2) \Theta(k^2-\p^2) ,
\label{Rkdef} 
\end{equation}
where $\Theta(x)$ is the step function. The $k$-dependent constant $Z_k$ is defined below.

The variation of the effective action with $k$ is given by Wetterich's equation~\cite{Wetterich93} 
\begin{equation}
\dk \Gamma_k[\phibf^\dagger,\phibf] = \half \Tr\llbrace \dk R_k\bigl(\Gamma^{(2)}_k[\phibf^\dagger,\phibf] + R_k\bigr)^{-1} \rrbrace , 
\label{rgeq} 
\end{equation} 
where $\Gamma_k^{(2)}[\phibf^\dagger,\phibf]$ denotes the second functional derivative of $\Gamma_k[\phibf^\dagger,\phibf]$. In Fourier space, the trace involves a sum over momenta, Matsubara frequencies and spin index.

\subsubsection{Scattering lengths $a_0$ and $a_2$}

In vacuum ($\mu=T=0$) the single-particle propagator is not renormalized. The renormalized interaction constant $g_{F,k}$ is simply obtained by summing the ladder diagrams. With the theta regulator~(\ref{Rkdef}), one finds 
\begin{equation}
\frac{1}{g_{F,k}} = \frac{1}{g_{F,\Lamb}} + \frac{M}{3\pi^2} (\Lamb-k) . 
\label{gFkvac} 
\end{equation}
This result can also be obtained from the RG equations~(\ref{rgeqQzeroT}). 
The scattering length $a_F\equiv a_F(g_{F,\Lamb},\Lamb)$ is then defined by 
\begin{equation}
g_{F,k=0} = \frac{4\pi a_F}{M} . 
\end{equation}
Note that in the model defined by~(\ref{ham5}), the cutoff $\Lamb$ cannot be arbitrarily large since stability of the system requires $g_{2,\Lamb}>0$ and $g_{0,\Lamb}+2g_{2,\Lamb}>0$. In the following we take $\Lamb=a_0^{-1}$ (with $a_0^{-1}<a_2^{-1}$ in all cases we shall consider). 

\subsubsection{Quantum limit $k\geq\Lamb_T$} 

Let us now turn to the calculation of the effective action $\Gamma_{\Lamb_T}$ at finite temperature and density ($T,\mu>0$). For $k\gtrsim\Lamb_T\sim\lamb^{-1}_{\rm dB}$ and $k\gtrsim \sqrt{2M\mu}$, the flow is dominated by the $T=\mu=0$ fixed point which controls the transition between the vacuum and the $T=0$ superfluid, and only quantum fluctuations contribute significantly to $\Gamma_k$. In three dimensions this fixed point is non-interacting so that all (properly defined) dimensionless many-body interactions are irrelevant and flow to zero. 

We assume the following form for the effective action,
\begin{equation}
\Gamma_k[\phibf^\dagger,\phibf] = \inttau \int d^3r \biggl\lbrace  \phibf^\dagger \left( \dtau - \frac{\nablabf^2}{2M} \right) \phibf 
+ U_k(\rho,\tau) \biggr\rbrace ,
\label{GamQ} 
\end{equation}
which corresponds to the LPA (with $Z_k=1$ in Eq.~(\ref{Rkdef})). 
For symmetry reasons, the effective potential $U_k(\rho,\tau)$ is a function of the O(3)$\times$U(1) invariants $\rho=\phibf^\dagger\phibf$ and $\tau=|\phibf\cdot\phibf|^2$. In the regime $k\geq\Lamb_T$ where the flow is dominated by the $\mu=T=0$ fixed point, it is natural to expand the effective potential in powers of $\rho$ and $\tau$,
\begin{equation}
U_k(\rho,\tau) = u_{0,k} + u_{1,k}\rho + \frac{u_{2,k}}{2} \rho^2 + v_{1,k}\tau , 
\label{Uvac}
\end{equation}
with 
\begin{equation}
\begin{gathered}
u_{0,\Lamb} = 0, \quad u_{1,\Lamb} = -\mu, \\
u_{2,\Lamb} = g_{2,\Lamb}, \quad v_{1,\Lamb} = \frac{g_{0,\Lamb}-g_{2,\Lamb}}{6} . 
\end{gathered} 
\end{equation}
The RG equations satisfied by $u_{0,k}$, $u_{1,k}$, $u_{2,k}$ and $v_{1,k}$ are given in Appendix~\ref{app_rgeq_quantum}. All dimensionless coupling constants $\tilde u_{i,k}=2Mk^{3i-5}u_{i,k}$ and $\tilde v_{i,k}=2Mk^{6i-5}v_{i,k}$ eventually flow to zero except $\tilde u_{0,k}$ and $\tilde u_{1,k}$~\footnote{The coupling constants $u_{i,k}$ and $v_{i,k}$ are associated to the terms $\rho^i$ and $\tau^i$ of the effective potential [Eq.~(\ref{Uvac})]. There are also (irrelevant) coupling constants $w_{ij,k}$ associated to $\rho^i\tau^j$ ($i,j>0$).}.

\begin{figure}
\centerline{\includegraphics[height=4cm]{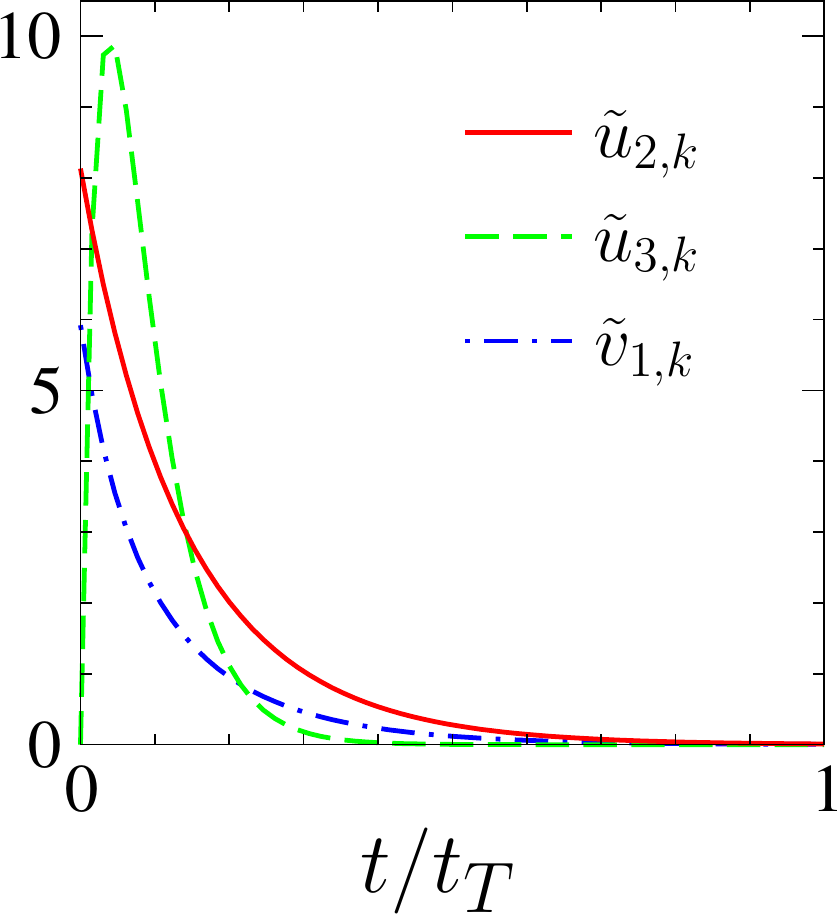}
\includegraphics[height=4cm]{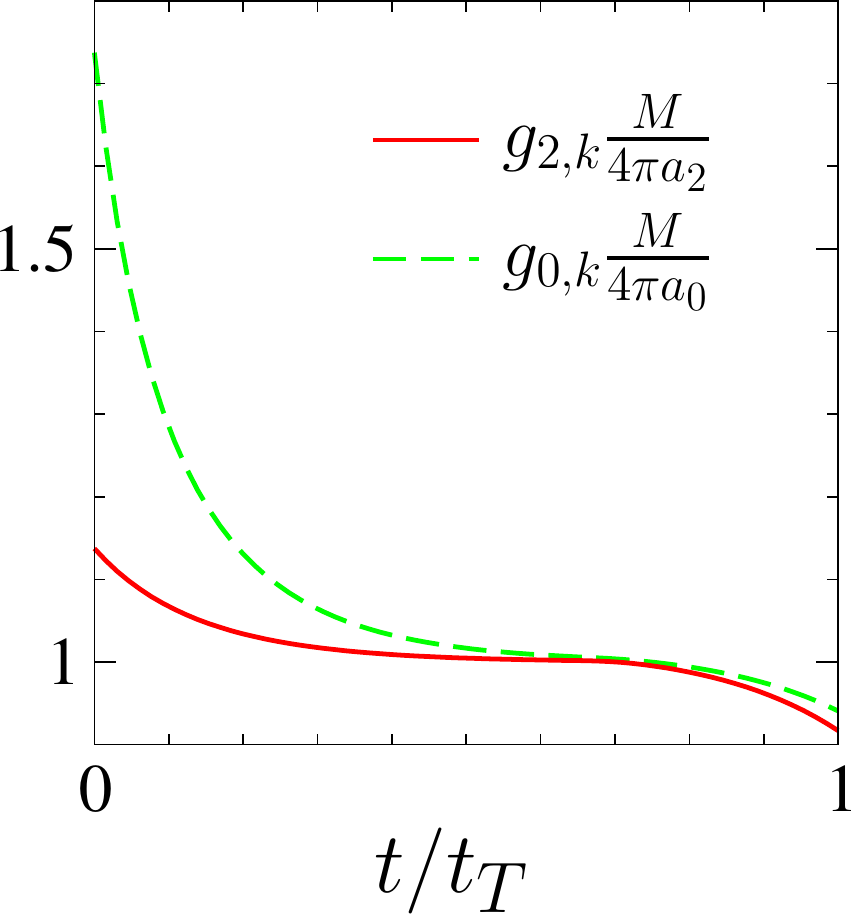}}
\caption{RG flow in the quantum limit $k\geq\Lamb_T$ with $\Lamb_T=\lamb^{-1}_{\rm dB}$ and initial conditions corresponding to $^{7}$Li with $Ma^2_F\mu\ll Ma_F^2T\ll 1$
[$t=\ln(k/\Lamb)$ and $t_T=\ln(\Lamb_T/\Lamb)$]. Left panel: Dimensionless coupling constants $\tilde v_{1,k}$, $\tilde u_{2,k}$ and $\tilde u_{3,k}$ (note that the latter is not included in the effective potential~(\ref{Uvac})). Right panel: $g_{0,k}$ and $g_{2,k}$ scaled by their $k=0$ values in vacuum.} 
\label{fig_rgquantum} 
\end{figure}

Figure~\ref{fig_rgquantum} shows the flow of some coupling constants with initial conditions corresponding to $^{7}$Li. We see that $g_{2,k}=u_{2,k}$ and $g_{0,k}=6v_{1,k}+u_{2,k}$, after a transient regime, become very close to their $k=0$ vacuum values $4\pi a_2/M$ and  $4\pi a_0/M$, respectively. When $k$ reaches $\lamb^{-1}_{\rm dB}$ thermal effects become noticeable, and $g_{2,k}=u_{2,k}$ and $g_{0,k}$ start deviating from their vacuum limit. 

By solving the RG equations to determine the precise value of the coupling constants at scale $\Lamb_T$ (rather than merely approximating $g_{F,\Lamb_T}=4\pi a_F/M$), we almost completely eliminate the dependence of the final results on the precise value of $\Lamb_T$ (see Sec.~\ref{subsec_num}). We have verified that including higher-order terms in~(\ref{Uvac}) leads to negligible changes in the final results discussed in Sec.~\ref{subsec_num}.

\subsubsection{Classical fluctuations: O(3)$\times$O(2) model}

The RG flow for $k\leq\Lamb_T$ is dominated by thermal fluctuations, i.e., fluctuations with vanishing Matsubara frequencies. To obtain the effective action $\Gamma_{k=0}$ from $\Gamma_{\Lamb_T}$, we therefore make a classical approximation where only fluctuations with $\wn=0$ are included. We have verified that the contribution to the flow of the first nonzero Matsubara frequency $\w_{n=1}=2\pi T$ is negligible (of the order of 0.02\%) compared to the contribution of $\w_{n=0}$.  
The NPRG approach simplifies in the classical O(3)$\times$O(2) model (no Matsubara sums have to be performed) and allows us to make a detailed study of the effective action, which is necessary to correctly describe the first order transition. As discussed in Sec.~\ref{sec_O3O2model} the (classical) Hamiltonian of the bosons is then given by the O(3)$\times$O(2) model [Eqs.~(\ref{ham4},\ref{ham2})].

To solve the RG equation~(\ref{rgeq}) for $k\leq\Lamb_T$ we use the so-called LPA$'$, an improvement of the LPA which includes a field renormalization factor $Z_k$~\cite{Berges02,Delamotte12}. In the language of Hamiltonian~(\ref{ham2}) we thus consider the effective action 
\begin{equation}
\Gamma_k[\phibf] = \int d^3r \biggl\lbrace \half Z_k [(\nablabf\phibf_1)^2+(\nablabf\phibf_2)^2] + U_k(\rho,\tau) \biggr\rbrace ,
\end{equation} 
where $\phibf_i=\mean{\varphibf_i}$ and 
$\rho$ and $\tau$ are defined in~(\ref{rtdef}) (with $\varphibf$ replaced by $\phibf$).
All the necessary information about the thermodynamics of the transition is included in the effective potential $U_k(\rho,\tau)$. For $k=\Lamb_T$, the latter is defined by 
\begin{equation}
U_{\Lamb_T}(\rho,\tau) = \beta u_{0,\Lamb_T} + r \rho + \frac{\lamb_1}{2} \rho^2 + \frac{\lamb_2}{2}\tau ,
\label{UMF} 
\end{equation}
where $r=2Mu_{1,\Lamb_T}$, $\lamb_1=(4M^2/\beta)u_{2,\Lamb_T}$ and $\lamb_2=(8M^2/\beta)v_{1,\Lamb_T}$. Without further integrating out fluctuations, the transition is predicted to be second order. 

Since $\tau$ vanishes in both the normal and superfluid phases, we perform a field semi-expansion of the effective potential,
\begin{equation}
U_k(\rho,\tau) = U_k^{(0)}(\rho) + \tau U_k^{(1)}(\rho) + \frac{\tau^2}{2} U_k^{(2)}(\rho) ,
\label{Uexpand}
\end{equation}
identical to that used in~\cite{Delamotte16} and which improves on the work reported in~\cite{Tissier00,Tissier03,Delamotte04}. 
A similar expansion has been used for a model with U($N$)$\times$U($N$) symmetry~\cite{Berges97a,Fukushima11,Fejos14}. Note that we make no expansion with respect to $\rho$. This allows the description of a first-order transition where a second local minimum may coexist with the minimum at $\rho=0$. Equation~(\ref{rgeq}) then yields four coupled equations for the three functions $U_k^{(i)}(\rho)$ ($i=0,1,2$) and the field renormalization factor $Z_k$. The flow equations are discussed in more detail in Appendix~\ref{app_rgeq_ONO2}.

\subsection{Numerical results} 
\label{subsec_num} 

For the numerical solution of the RG equations, we use the known values of $a_0$ and $a_2$ for the Bose gas of interest ($^{87}$Rb, $^{41}$K or $^7$Li) and choose a typical experimental value for the density $n$~\footnote{We choose a value of the density corresponding to a recent experiment where the atoms were trapped in a quasi-uniform potential~\cite{Gaunt13}.}. We set the temperature equal to the BEC temperature $T_c^0=(2\pi/M)(n/3\zeta(3/2))^{2/3}$ of the noninteracting gas, and choose $\Lamb_T=\lamb^{-1}_{\rm dB}$ (we shall see that the precise value of $\Lamb_T$ does not significantly affect the final results). 
To locate the transition, we vary the chemical potential $\mu$ and look for the absolute minimum of the effective potential $U^{(0)}_k(\rho)$ in the limit $k\to 0$. The superfluid (ferromagnetic) phase corresponds to a nonzero value of the position $\rho_0$ of the absolute minimum of the effective potential~\footnote{The actual value of the density at the transition differs from the value used to determine the temperature. We discuss this issue in Sec.~\ref{subsec_pseudoscaling}.}. 

\subsubsection{$^{87}${\rm Rb} and $^{41}${\rm K}} 

\begin{figure}
\centerline{\includegraphics[width=6cm]{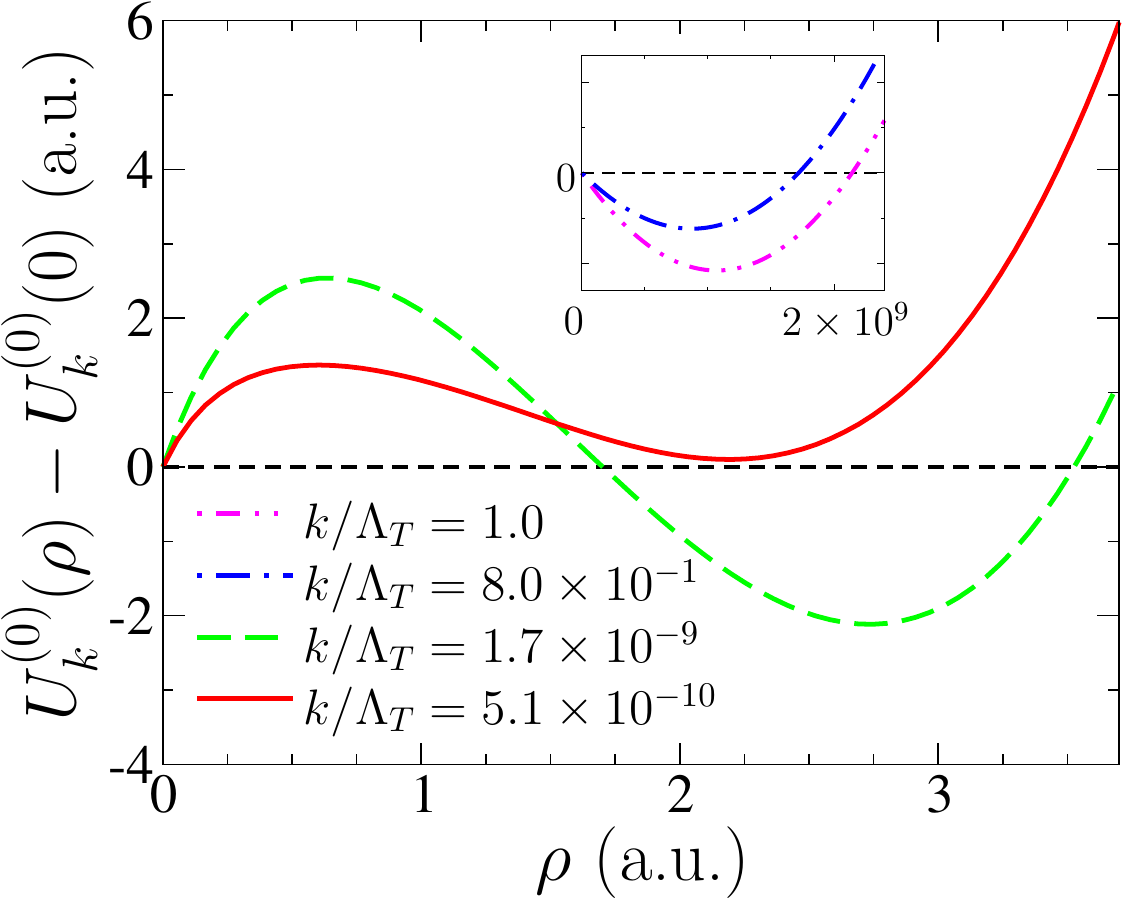}}
\caption{Effective potential $U^{(0)}_k(\rho)$ vs $\rho$ for various values of $k$ and $\mu=\mu_c$ with initial conditions at $k=\Lamb_T$ corresponding to $^{87}$Rb. The potential exhibits a single minimum at the beginning of the RG flow (see inset) whereas 2 minima coexist for sufficiently small $k$.}
\label{fig_U0k}
\centerline{\includegraphics[width=6.5cm]{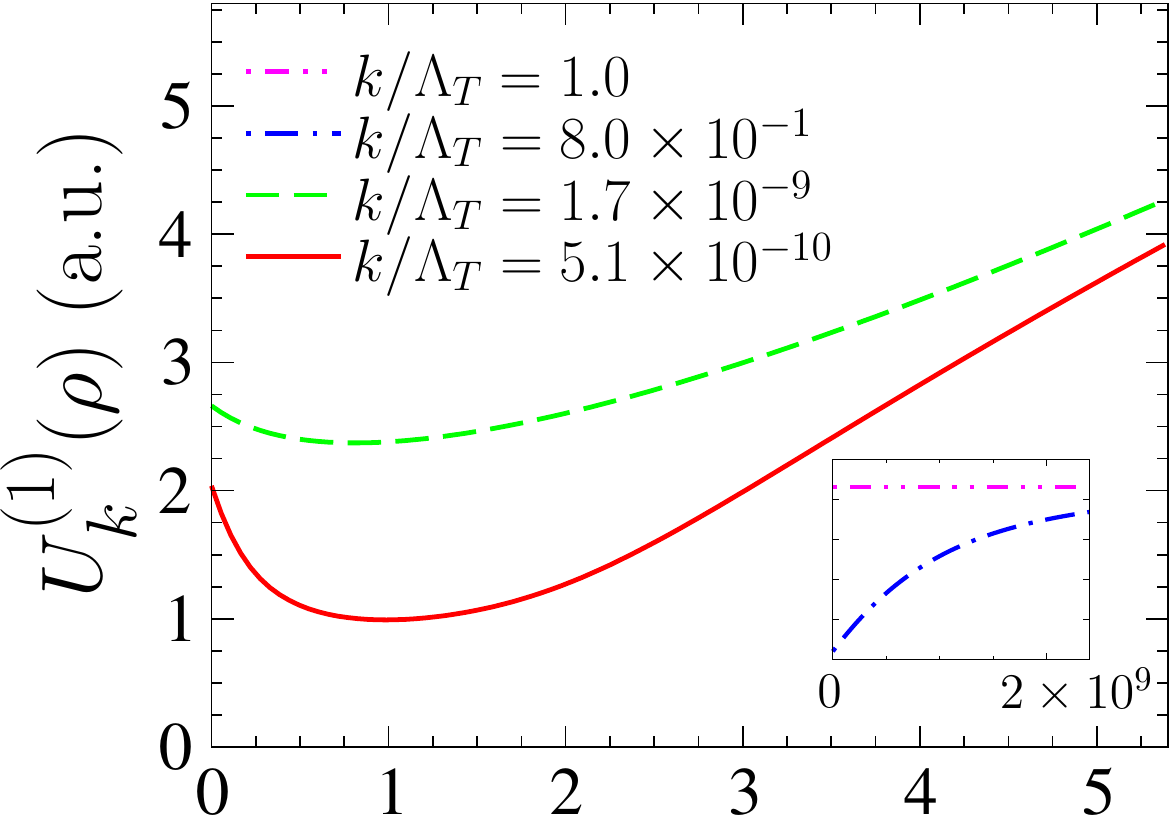}}
\vspace{0.2cm}
\centerline{\includegraphics[width=6.5cm]{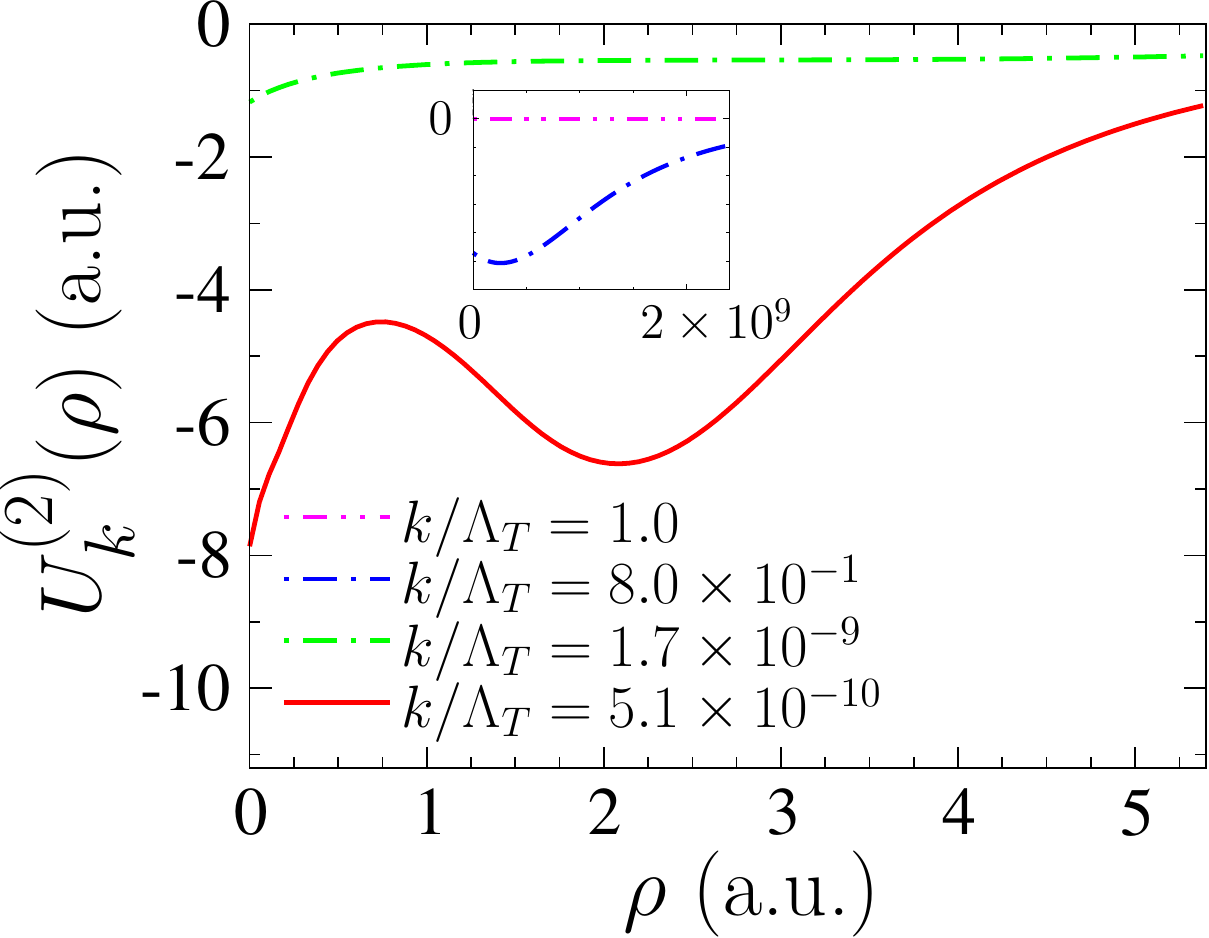}}
\caption{Effective potentials $U^{(1)}_k(\rho)$ and $U^{(2)}_k(\rho)$ vs $\rho$ for various values of $k$ and $\mu=\mu_c$ ($^{87}$Rb).}
\label{fig_U1k}
\end{figure}

Let us first discuss the case of a $^{87}$Rb atom gas. 
Figure~\ref{fig_U0k} shows the $k$-dependence of the effective potential $U_k^{(0)}(\rho)$ at the transition ($\mu=\mu_c$). Initially, for $k=\Lamb_T$, the system is ordered and $U_{\Lamb_T}^{(0)}(\rho)=r\rho+(\lamb_1/2)\rho^2$ shows a minimum at a nonzero value $\rho_{0,\Lamb_T}=-r/\lamb_1$. The effect of fluctuations is twofold. Long-range order is suppressed as $k$ decreases (i.e., $\rho_{0,k}$ decreases) and for sufficiently small $k$ a second minimum appears at $\rho=0$.  Both minima become degenerate when $k\to 0$. For $\mu<\mu_c$, the minimum at $\rho=0$ is the absolute minimum (normal phase), whereas the nontrivial minimum is the absolute one when $\mu>\mu_c$ (superfluid phase). As a consequence the order parameter makes a discontinuous jump at the phase transition (implying a discontinuous jump $\Delta n_0$ of the condensate density), which is therefore (fluctuation-induced) first order. The potentials $U^{(1)}_k(\rho)$ and $U^{(2)}_k(\rho)$ are shown in Fig.~\ref{fig_U1k} (with initial conditions $U^{(1)}_{\Lamb_T}(\rho)=\lamb_2/2$ and $U^{(2)}_{\Lamb_T}(\rho)=0$). The first-order phase transition in spin-one Bose gases has also been inferred from a two-loop RG approach to the classical Hamiltonian~(\ref{ham3}) in $d=4-\eps$ dimensions~\cite{Szirmai06,note20}.

The RG equation $\dk U_k$ is unstable for small $k$ so that it is not possible to determine the effective potential for arbitrary small values of $k$. This instability is due to a pole appearing in the propagator at a finite value $k_c$ of the RG momentum scale $k$, which prevents continuing the flow for $k<k_c$~\footnote{The pole is due to the propagator becoming negative when $\rho$ is near the local maximum located between the two minima of $U_k^{(0)}$.}. Similar instabilities have been encountered in previous studies of first-order transitions~\cite{Berges97a,note21}. 
Nevertheless, we find that all physical quantities of interest (e.g. the location of the minima of $U_k^{(0)}(\rho)$ or the correlation length) have nearly converged before the instability occurs~\footnote{In practice, we extrapolate the results to $k=0$ in order to improve our estimates.}. 

\begin{figure}
\centerline{\includegraphics[width=7cm]{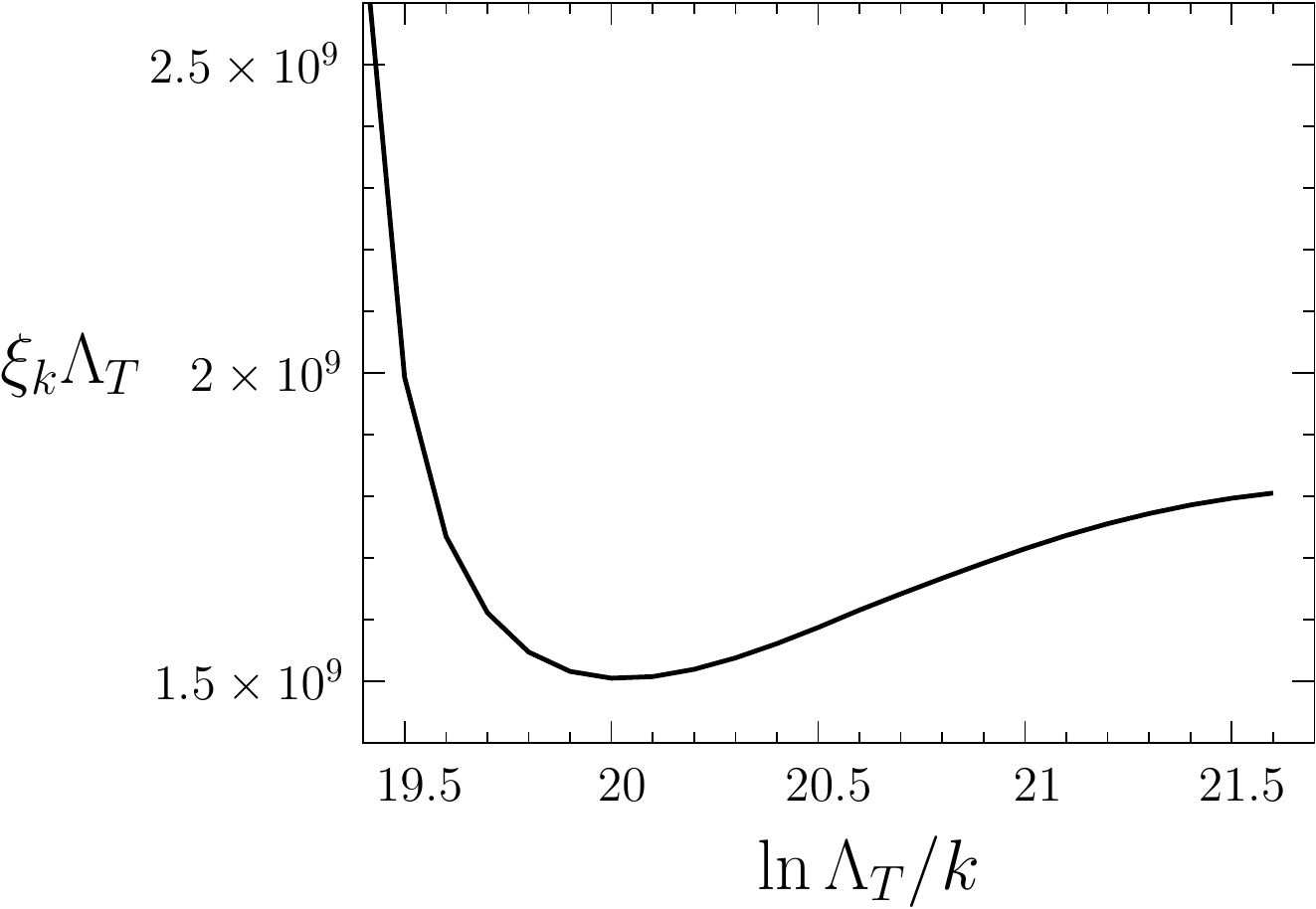}}
\caption{$k$-dependent correlation length $\xi_k$ [Eq.~(\ref{xidef})] vs $\ln(\Lamb_T/k)$ for initial conditions corresponding to $^{87}$Rb.}
\label{fig_xicv} 
\end{figure}

The correlation length is defined by
\begin{equation}
\xi \equiv \lim_{k\to 0} \xi_k = \lim_{k\to 0}\left( \frac{Z_k}{U_k^{(0)}{}'(0)} \right)^{1/2} 
\label{xidef} 
\end{equation}
(see Appendix~\ref{app_rgeq_ONO2}). At the transition (or at temperatures infinitesimally above $T_c$), where $U^{(0)}_{k=0}(0)=U^{(0)}_{k=0}(\rho_0)$, convexity of the effective potential implies that $U_{k=0}^{(0)}(\rho)$ must be constant for $0\leq\rho\leq\rho_0$. Here $\rho_0=\lim_{k\to 0}\rho_{0,k}$ denotes the position of the nontrivial minimum in the limit $k\to 0$. To reconcile this observation with a finite value of the correlation length, one must assume that for any nonzero $k$ there is a region around $\rho=0$, whose size vanishes when $k\to 0$, where the effective potential shows a nonzero derivative, i.e., $U_k^{(0)}{}'(0)>0$. This is not in contradiction with convexity requirement since the effective potential needs to be convex only when $k=0$~\footnote{This is due to the scale-dependent effective action $\Gamma_k[\phibf]$ being defined as a slightly modified Legendre transform [Eq.~(\ref{Gamdef})], so that the true (convex) effective potential is $U_k(\rho,\tau)+\rho R_k(\p=0)$, which coincides with $U_k(\rho,\tau)$ only for $k=0$.}. 
This scenario seems to agree with our numerical results. Although the flow cannot be continued down to $k=0$ due to the numerical instabilities mentioned above, we find that $\xi_k$ converges towards a nonzero value which we identify with the correlation length~(\ref{xidef}) (Fig.~\ref{fig_xicv}). 

\begin{figure}
\centerline{\includegraphics[height=3.75cm]{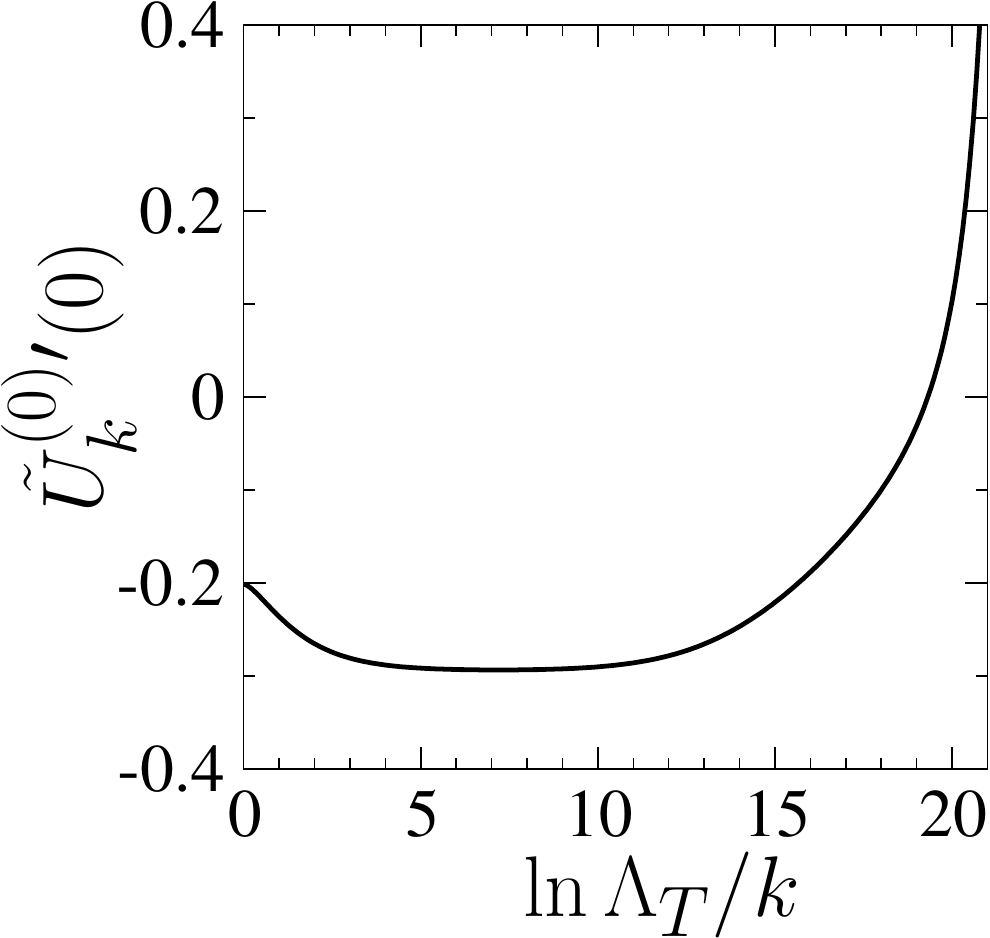}
\includegraphics[height=3.75cm]{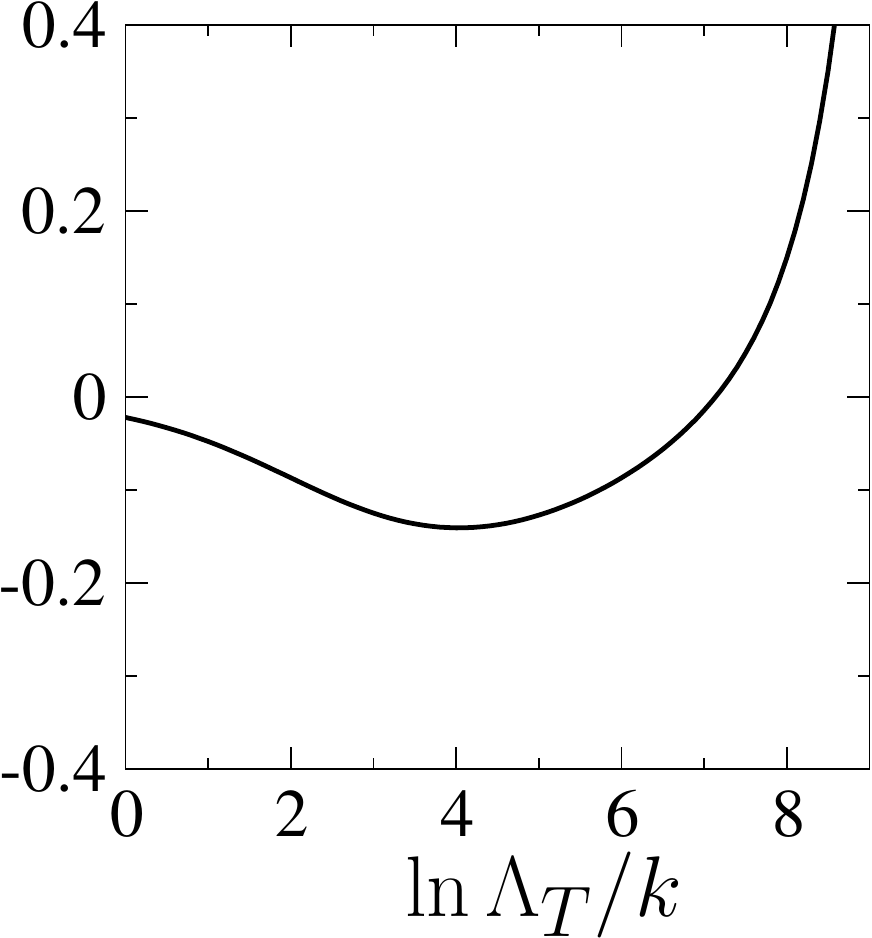}}
\caption{Left panel: $\tilde U^{(0)}_k{}'(0)$ vs $k$ for $\mu\simeq\mu_c$ ($^{87}$Rb). For $5\lesssim k\lesssim 12$, one observes a quasi-plateau due to the proximity of the RG trajectory to the O(6) Wilson-Fisher fixed point. Right panel: same as left panel but for $^7$Li.}
\label{fig_U0tildep}  
\centerline{\includegraphics[width=6.5cm]{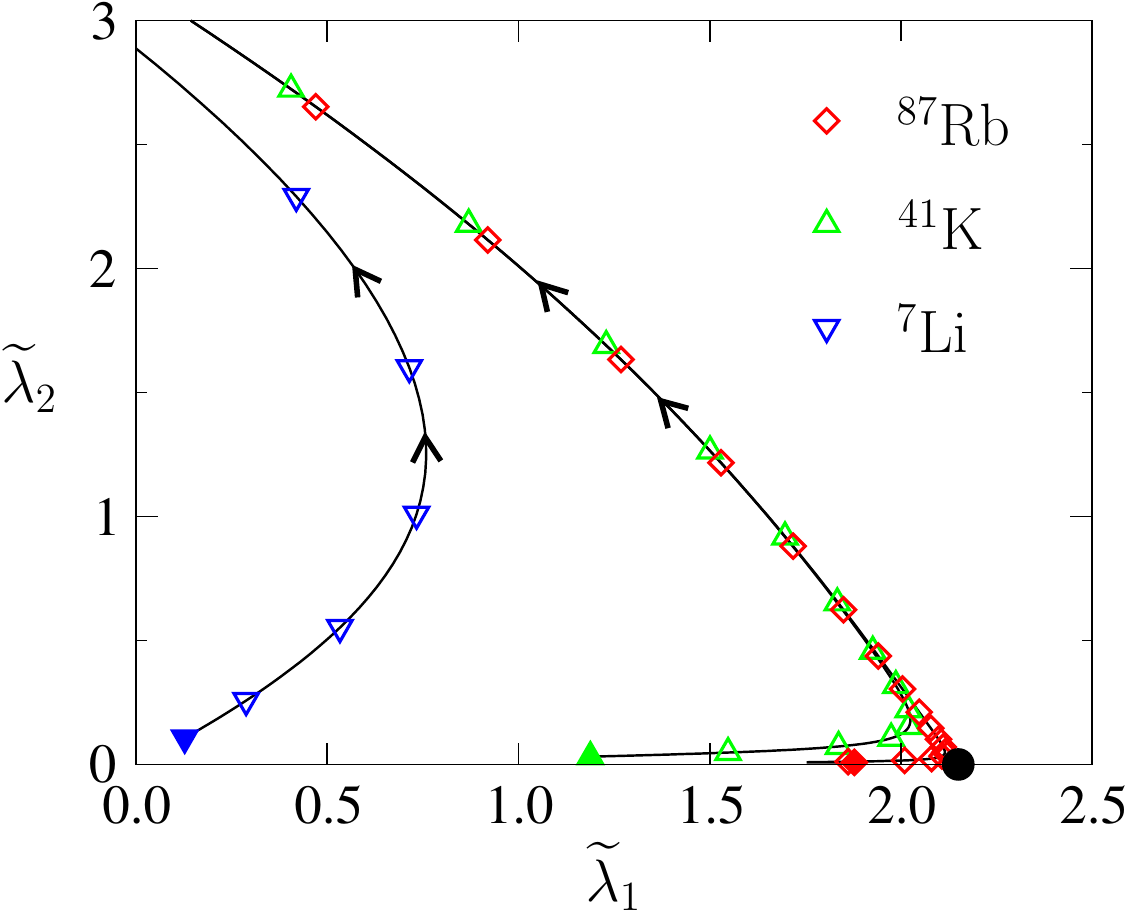}}
\caption{Critical RG flow diagram in the plane $(\tilde\lamb_{1},\tilde\lamb_{2})$. The symbols correspond to equal steps in $\ln(\Lamb_T/k)$ (solid ones show the initial conditions of the trajectories). The black dot shows the Wilson-Fisher fixed point of the three-dimensional O(6) model.
}
\label{fig_flow}
\end{figure}

At the transition the correlation length of $^{87}$Rb is several orders of magnitude larger than $\lamb_{\rm dB}$, i.e., much larger than the size $L\sim 40\lamb_{\rm dB}$ of the system in a typical experiment: the transition is weakly first order. This implies that neither the finiteness of $\xi(\mu_c)$ nor the jump $\Delta n_0\sim n\lamb_{\rm dB}/\xi$ [Eq.~(\ref{dn0})] of the condensate density can be observed experimentally (for a summary of results for $^{87}$Rb, $^{41}$K and $^7$Li see Table~\ref{table_nu} and Fig.~3 in Ref.~\cite{Debelhoir16a}).

The extremely large correlation length in $^{87}$Rb can be understood by considering the flow of the dimensionless effective potential $\tilde U^{(0)}_k(\trho)=k^{-3} U^{(0)}_k(\rho)$ where $\trho=Z_k k^{-1}\rho$. If the transition were second-order, $\tilde U^{(0)}_k(\trho)$ would reach a ($k$-independent)  fixed-point value $\tilde U^{(0)*}(\trho)$ at the critical point. Figure~\ref{fig_U0tildep} shows $\tilde U_k^{(0)}{}'(0)$ vs $k$ for $\mu\simeq\mu_c$. After a transient regime, $\tilde U_k^{(0)}{}'(0)$ reaches a quasi-plateau where it is nearly $k$-independent, before the flow eventually runs away. The origin of this behavior appears clearly if one considers the RG trajectory projected onto the plane ($\tlamb_{1,k}$,$\tlamb_{2,k}$) where the dimensionless coupling constant $\tlamb_{i,k}=\lamb_{i,k}/Z_k^2 k$ is defined by 
\begin{equation}
\lamb_{1,k} = U_k^{(0)}{}''(0), \quad \lamb_{2,k} = 2 U_k^{(1)}{}(0) 
\end{equation}
(see Eq.~(\ref{UMF})). For initial conditions corresponding to $^{87}$Rb, the RG trajectory is strongly drawn to the vicinity of the O(6) Wilson-Fisher fixed point, where the flow is very slow and all (properly defined) dimensionless quantities remain nearly constant as $k$ varies (Fig.~\ref{fig_flow}). The long ``time'' spent in the vicinity of the O(6) Wilson-Fisher fixed point explains the very large value of the correlation length at the transition.

$^{41}$K does not differ noticeably from $^{87}$Rb, the scattering lengths $a_0$ and $a_2$ having similar values for both types of atoms. Thus we find that the RG trajectories for initial conditions corresponding to $^{41}$K and $^{87}$Rb are similar. In both cases, the flow is strongly influenced by the O(6) Wilson-Fisher fixed point (Fig.~\ref{fig_flow}) and the correlation length $\xi(\mu_c)$ at the transition is extremely large (Table~\ref{table_nu}). The flow eventually runs away, as expected for a first-order phase transition. 

\begin{figure}
\centerline{\includegraphics[width=6cm]{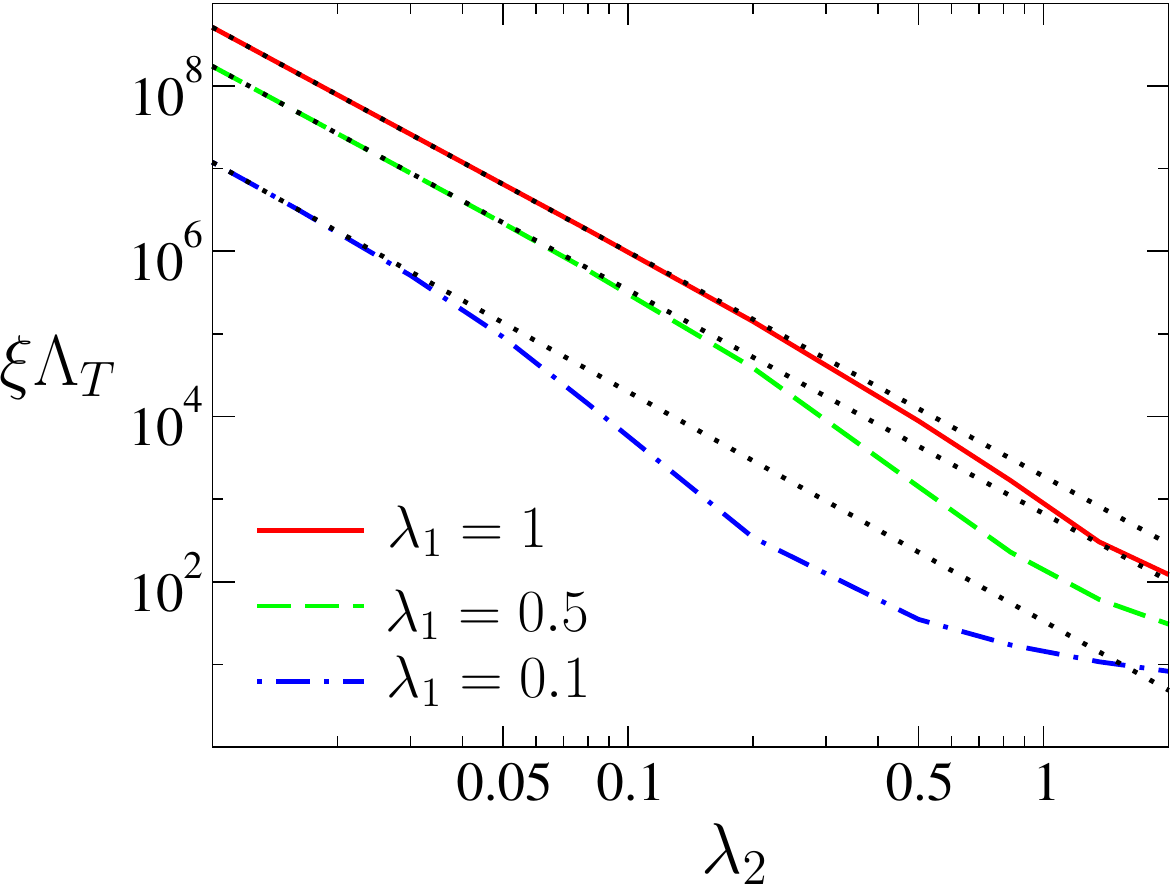}}
\caption{Correlation length $\xi$ at the transition as a function of $\lamb_2$ for various values of $\lamb_1$. The dotted lines correspond to $\xi\sim\lamb_2^{-1/y}$ with $y\simeq 0.367$.}
\label{fig_xi_lamb2}
\end{figure}

Whenever $\lamb_2$ is small (as in the case of $^{87}$Rb or $^{41}$K), the RG trajectories spend a long RG time in the vicinity of the Wilson-Fisher O(6) fixed point before eventually running away. As a result, the correlation length $\xi\sim \lamb_2^{-1/y}$ is determined by the inverse of the eigenvalue $y\simeq 0.367$ of the linearized flow equations corresponding to the unstable direction of the O(6) fixed point (the other positive eigenvalue gives the inverse of the correlation-length exponent $\nu_{\rm O(6)}$) (Fig.~\ref{fig_xi_lamb2}). Deviations from the behavior $\xi\sim \lamb_2^{-1/y}$ become significant when $\lamb_2$ is at least of the order of $\lamb_1$. Similar results have been obtained in a U($N$)$\times$U($N$) model~\cite{Berges97a}.

\subsubsection{$^7${\rm Li}} 

The case of $^7$Li is significantly different. The quasi-plateau observed in the flow of $\tilde U^{(0)}_k{}'(0)$ for $^{87}$Rb has essentially disappeared (Fig.~\ref{fig_U0tildep}, right panel) and the RG trajectory is clearly not significantly influenced by the O(6) Wilson-Fisher fixed point (Fig.~\ref{fig_flow}). As a consequence, the correlation length $\xi(\mu_c)$ at the transition is much shorter although still much larger than the size of the gas in a typical experiment (Table~\ref{table_nu}).

\subsection{Pseudoscaling} 
\label{subsec_pseudoscaling} 

\begin{table}
\caption{Correlation length $\xi(\mu_c)$, condensate-density jump $\Delta n_0$ and pseudocritical exponent $\nu$. The values of $a_0$ and $a_2$ are taken from Ref.~\cite{Stamper-Kurn13}. $a_B$ denotes the Bohr radius and $\lamb_{\rm dB}$ the thermal de Broglie wavelength.}
\label{table_nu} 
\begin{tabular}{cccc}
\hline\hline
 & $^{87}$Rb &$^{41}$K & $^7$Li
\\ \hline 
$a_0/a_B$ & $101.8\pm 0.2$ & $68.5\pm 0.7$ & 23.9
\\ 
$a_2/a_B$ & $100.4\pm 0.1$ & $63.5\pm 0.6$ & 6.8 
\\
$\xi(\mu_c)/\lamb_{\rm dB}$ & $2.4\times 10^9$ & $3.6\times 10^7$ & $8.7\times 10^3$ 
\\
$\Delta n_0 \lamb^3_{\rm dB}$ & $2.1\times 10^{-9}$ & $1.6\times 10^{-7}$ & $1.0\times 10^{-3}$  
\\ 
$\nu$ & 0.78 & 0.77  & 0.60
\\ \hline \hline
\end{tabular} 
\end{table}

As first pointed out in the context of the magnetic transition in STHAs~\cite{Zumbach93}, the strong increase of the correlation length as the transition is approached allows one to define a (nonuniversal) pseudocritical exponent $\nu$ by $\xi\sim (\mu_c-\mu)^{-\nu}$: $\ln\xi$ depends quasi-linearly on $\ln(\mu_c-\mu)$ on several decades with a slope which varies typically by a few percents: 5\% for $^{87}$Rb and 2\% for $^7$Li. We define the exponent $\nu$ by the value of the slope when $\xi$ is of the order of the size $L\sim 40\lamb_{\rm dB}$ of the system. 
Note that the same exponent $\nu$ characterizes the increase of the correlation length, i.e., $\xi\sim(T-T_c)^{-\nu}$, when the transition is approached at fixed chemical potential by varying the temperature. The value of $\nu$ for $^{87}$Rb, $^{41}$K and $^7$Li is reported in Table~\ref{table_nu}. 

The exponent $\nu$ varies by less than 0.02\% if we include only $U_k^{(0)}(\rho)$ and $U_k^{(1)}(\rho)$ in~(\ref{Uexpand}), which shows that the field semi-expansion has nearly converged. Furthermore, higher-order derivative terms not included in the LPA$'$ are expected to be essentially irrelevant for the computation of $\nu$ when, as is the case here, the running anomalous dimension $\eta_k=-k\dk\ln Z_k\lesssim 0.08$ is small. We have also studied the dependence of $\nu$ on the choice of the  infrared regulator. With the exponential regulator $R_k(\p)=\alpha \p^2/(e^{\p^2/k^2}-1)$  
and choosing $\alpha$ following the principle of minimum sensitivity~\footnote{The principle of minimum sensitivity stipulates that the best estimate of $\nu$ is stationary with respect to change in $\alpha$, i.e., $d\nu/d\alpha=0$.}, we find $\nu\simeq 0.59$ for $^7$Li, in good agreement with the results obtained with the theta regulator~(\ref{Rkdef}).
We have also verified that our results are independent of the choice of the momentum cutoff $\Lamb_T$. Varying $\Lamb_T$ between $0.25\lamb^{-1}_{\rm dB}$ and $4\lamb^{-1}_{\rm dB}$, we find that $\xi/\lamb_{\rm dB}$ and $\Delta n_0\lamb^3_{\rm dB}$ change by 0.5\% and $\nu$ by 0.3\%. 

Finally, we have verified that the pseudocritical exponent $\nu$ is essentially independent of the transition temperature, i.e., the density (or, equivalently, the chemical potential). For $^7$Li, $\nu$ is equal to 0.61 and 0.59 if we multiply the density by 5 and 0.2, respectively~\footnote{The value of $\nu$ quoted here corresponds to a fixed total number of atoms (if we multiply the density by a factor 5, we multiply the size $L$ of the system by $5^{1/3}$).}. So far we have loosely related the density to the transition temperature using the expression $T_c^0=(2\pi/M)(n/3\zeta(3/2))^{2/3}$ of the BEC temperature of the noninteracting spin-one Bose gas. A more precise relation can be obtained using $n=\partial P(\mu,T)/\partial\mu$ where $P(\mu,T)=-U_{k=0}(\rho_{0,k=0},\tau=0)$ is the pressure. The density is discontinuous at the transition as expected for a first-order transition. For instance, in the case of $^7$Li, $\Delta n/n\sim 5\times 10^{-3}$ and is of the same order as the jump of the condensate density $\Delta n_0/n\sim \Delta n_0\lamb^3_{\rm dB}\sim 10^{-3}$ (Table~\ref{table_nu}). The critical density ($\mu\to\mu_c^\pm$ or $T\to T_c^\mp$) differs by less than 10\% from the noninteracting result $n^0_c=3\zeta(3/2)(MT/2\pi)^{3/2}$. Similarly we find that the critical value $\mu_c$ of the chemical potential differs by typically 5\% from the Hartree-Fock result
\begin{equation}
\mu_c^{\rm HF} = \frac{8\pi\zeta(3/2)(a_0+5a_2)}{3M} \left( \frac{MT}{2\pi} \right)^{3/2} . 
\label{muHF}
\end{equation}

We find that the regime where pseudoscaling holds is reached as soon as $\xi$ becomes larger than the thermal de Broglie wavelength $\lamb_{\rm dB}$, which suggests that the Ginzburg length $\xi_G$ is of the order of $\lamb_{\rm dB}$. The Ginzburg criterion predicts
\begin{equation}
\frac{\xi_G}{\lamb_{\rm dB}} = \frac{\alpha}{\sqrt{Ma^2T}} 
\end{equation}
for a spin-zero boson gas, with $\alpha$ a constant. Using the results of Ref.~\cite{Giorgini96}, one finds $\alpha\sim 10^{-2}$. Although the Ginzburg criterion is too crude to give a reliable value of $\alpha$, the small value found here suggests that the ratio $\xi_G/\lamb_{\rm dB}$ can be close to one even for small values of $Ma^2T$. Our numerical results correspond to $Ma^2_0T\sim Ma^2_2T\sim 10^{-4}$ for $^{87}$Rb, $Ma^2_0T\sim 10^{-5}$ and $Ma^2_2T\sim 10^{-6}$ for $^7$Li, and are compatible with $\xi_G\sim\lamb_{\rm dB}$. 

The value of $\nu$ in $^{87}$Rb and $^{41}$K atom gases, which is close to the value $\nu_{\rm O(6)}\simeq 0.83$, is largely a consequence of a crossover phenomenon due to the proximity of the O(6) Wilson-Fisher fixed point, and is independent of the ultimate first-order character of the transition. By contrast, the value $\nu\simeq 0.60$ in $^7$Li is not related in any way to the existence of a nearby critical fixed point: this value is {\it nonuniversal} and depends solely on the scattering lengths $a_0$ and $a_2$. In Sec.~\ref{sec_ONO2}, we show that pseudoscaling in $^7$Li is actually linked the (true) scaling behavior in the O($N$)$\times$O(2) model when $N>N_c\simeq 5.3$. 

\subsection{Ferromagnetic transition without superfluidity}
\label{subsec_ferrowosf} 

In Ref.~\cite{Natu11} it has been predicted that the (assumed second-order) superfluid transition to the ferromagnetic phase becomes a ferromagnetic transition without BEC when $a_0/a_2>5/2$, this latter condition being verified in the $^7$Li atom gas. This result has been obtained by comparing the BEC temperature $T_c^0$ of the noninteracting spin-one Bose gas with the transition temperature $T_f$ of the ferromagnetic transition (without BEC) obtained in the random-phase approximation. Momentarily discarding the first-order character of the superfluid transition, we note that the condition $a_0/a_2>5/2$ is likely to be highly sensitive to fluctuations since the difference between $T_c^0$ and $T_f$ is of order $n^{1/3}a_F$. For instance, fluctuations are known to increase the BEC temperature in a spin-zero Bose gas by a quantity of order $n^{1/3}a$: $\Delta T_c/T_c^0\simeq 1.32 n^{1/3}a$~\cite{Arnold01}. Assuming that the increase of $T_c^0$ in a spin-one Bose gas is given by $\Delta T_c/T^0_c\simeq 1.32[(a_0+5a_2)/3](n/3)^{1/3}$~\footnote{The Hartree-Fock self-energy of a spin-one Bose gas suggests that the scattering length and the density entering the formula $\Delta T_c/T_c^0\simeq 1.32 n^{1/3}a$ are $(a_0+5a_2)/3$ and $n/3$ respectively.} and using the results of~\cite{Natu11} for $T_f$, we find that $T_f>T_c$ if 
\begin{equation}
\alpha\left(\third+\frac{a_2-a_0}{a_0+2a_2}\right) (a_0+2a_2) + \frac{\beta}{3^{4/3}}(a_0+5a_2) < 0 , 
\label{critferro}
\end{equation}
where $\alpha\simeq 1.61$ and $\beta\simeq 1.32$ (see Appendix~\ref{sec_ferro}). $^7$Li does not satisfy condition~(\ref{critferro}) so that $T_c>T_f$: the normal phase is unstable against a superfluid (ferromagnetic) transition. On the other hand fluctuations are expected to decrease $T_f$ and therefore suppress the ferromagnetic transition even further. 

Since the superfluid transition is likely to first order, as shown in this paper, the possible existence of a ferromagnetic transition without BEC remains an open question.

\section{Scaling and pseudoscaling in the O($N$)$\times$O(2) model}
\label{sec_ONO2} 

\begin{figure}
\centerline{\includegraphics[angle=0,width=8cm]{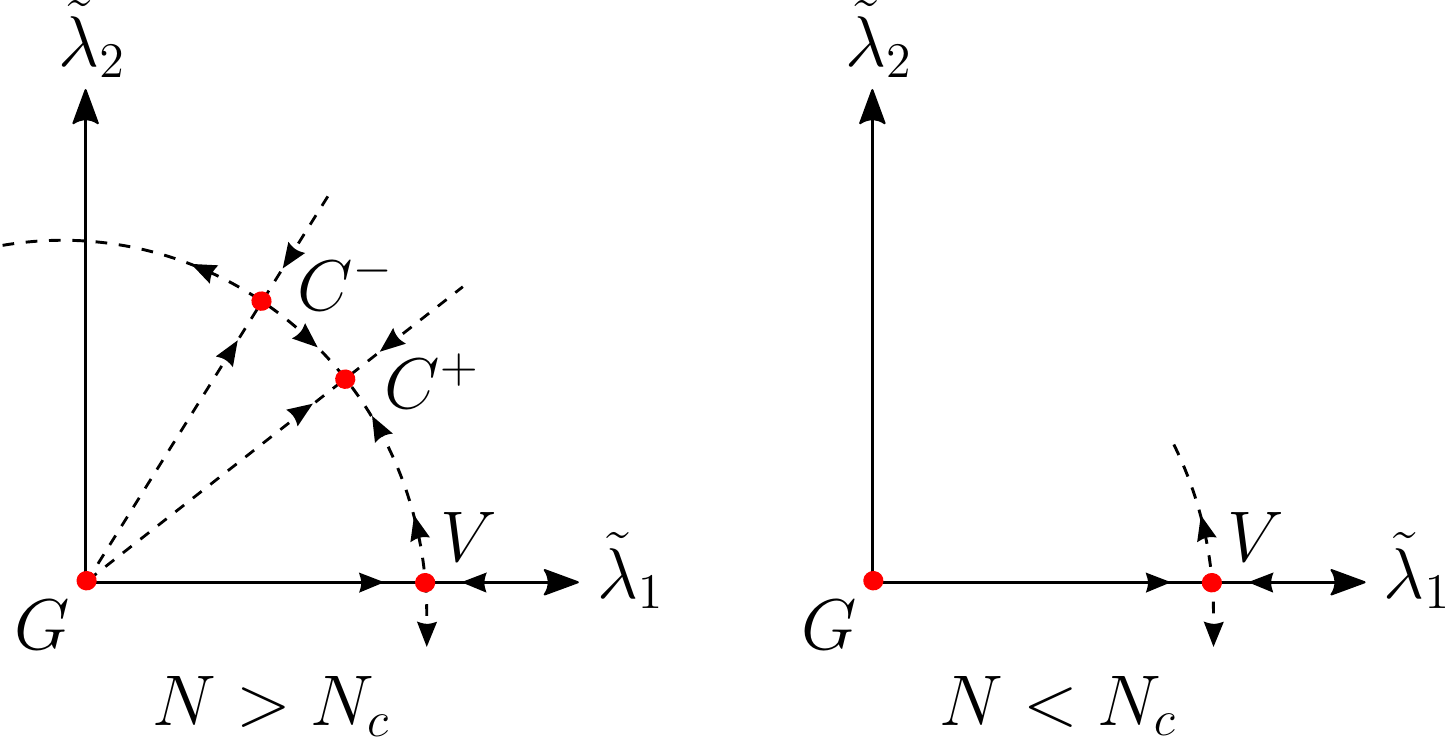}}
\caption{Schematic flow diagram of the three-dimensional O($N$)$\times$O(2) model according to perturbative RG near four dimensions and NPRG. $C^+$ and $C^-$ denote the chiral and antichiral fixed points, respectively, V the O($2N$) Wilson-Fisher fixed points and $G$ the Gaussian fixed point. The chiral and antichiral fixed points merge when $N=N_c$ and no stable fixed points are present when $N<N_c$.} 
\label{fig_flowONO2}   
\centerline{\includegraphics[height=4.6cm]{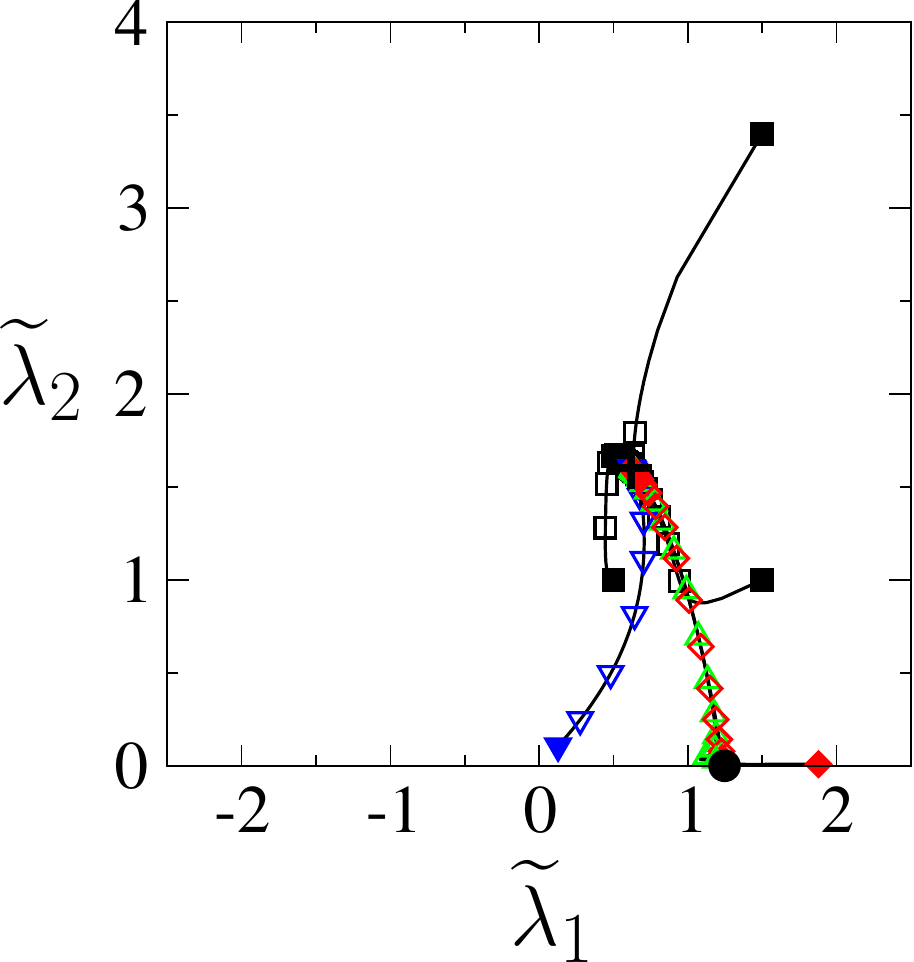}
\includegraphics[height=4.6cm]{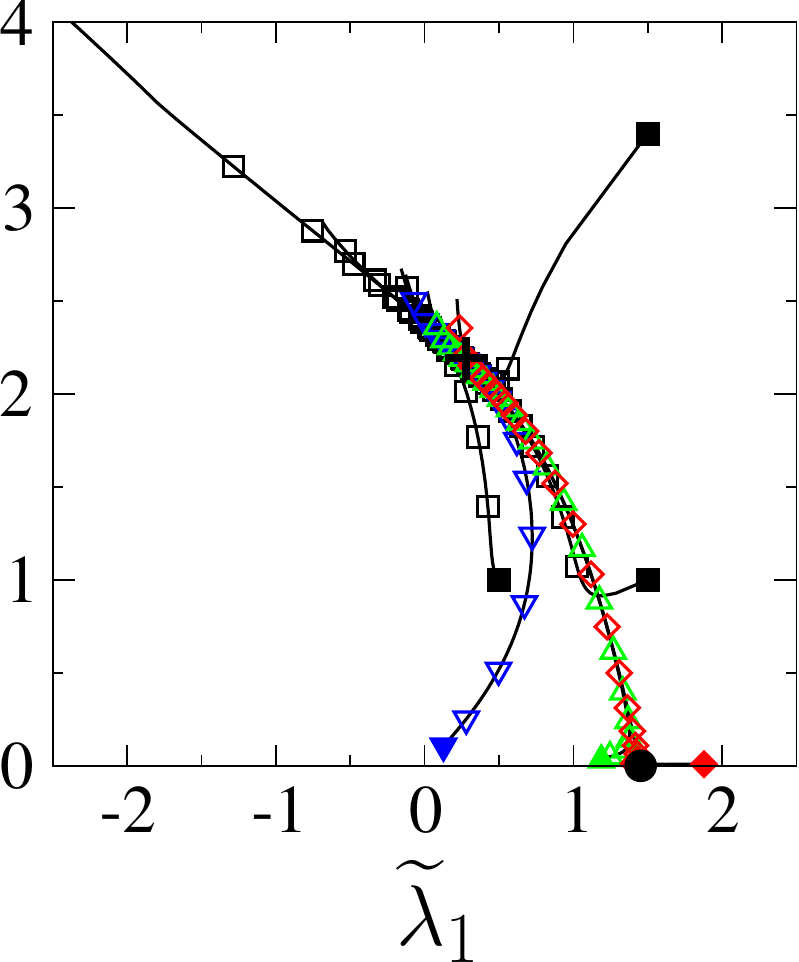}}
\caption{Left panel: critical flow diagram of the O(6)$\times$O(2) model in the plane ($\tlamb_{1},\tlamb_{2}$) as obtained from the NPRG. For each trajectory the symbols correspond to equal steps in $t=\ln(k/\Lamb_T)$ (initial conditions are shown by a solid symbol). All trajectories are drawn to the chiral fixed point ($+$ symbol). Trajectories shown by red ($\textcolor{red}{\Diamond}$), green ($\textcolor{green}{\triangle}$) and blue ($\textcolor{blue}{\triangledown}$) symbols correspond to $^{87}$Rb, $^{41}$K and $^7$Li, respectively. The black dot shows the O(12) Wilson-Fisher fixed point. 
Right panel: same as left panel but for the O(5)$\times$O(2) model. The $+$ symbol shows the projection onto the plane $(\tlamb_1,\tlamb_2)$ of the two unphysical fixed points with complex coordinates. The black dot shows the O(10) Wilson-Fisher fixed point.}
\label{fig_N65}
\centerline{\includegraphics[height=4.6cm]{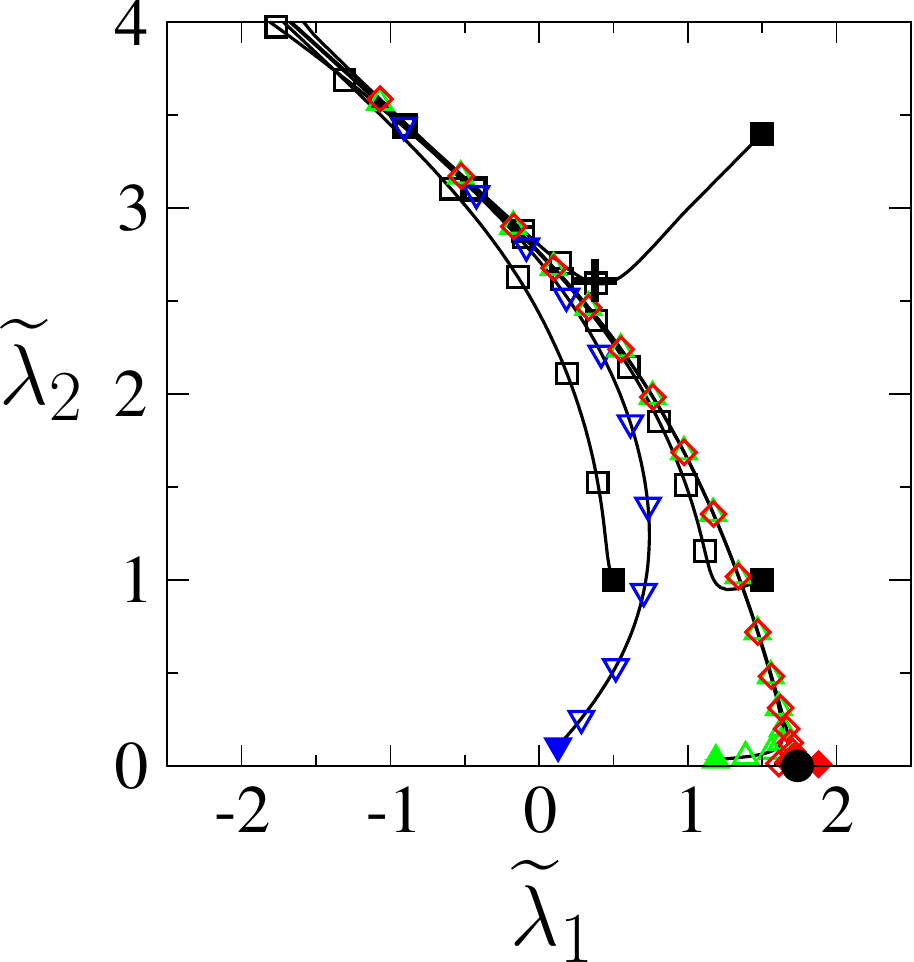}
\includegraphics[height=4.6cm]{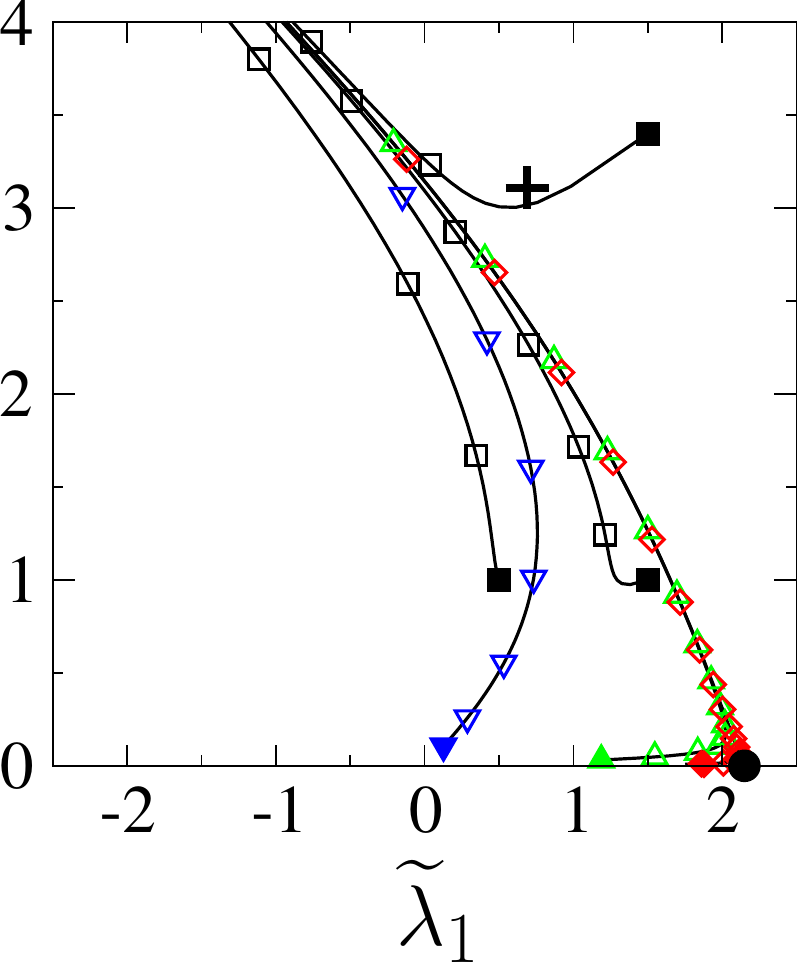}}
\caption{Same as Fig.~\ref{fig_N65} but for the O(4)$\times$O(2) (left) and O(3)$\times$O(2) (right) models.}
\label{fig_N34}
\end{figure}

Figure~\ref{fig_flowONO2} shows the schematic flow diagram of the three-dimensional O($N$)$\times$O(2) model according to perturbative RG near four dimensions~\cite{Garel76,Bailin77,Yosefin85,Antonenko95a,Holovatch04,Calabrese04} and NPRG~\cite{Tissier00,Tissier03,Delamotte04,Delamotte16}. Above a critical value $N_c$, the transition is second order and governed by a (stable) ``chiral'' fixed point (denoted by $C^+$ in Fig.~\ref{fig_flowONO2}) which coexists with three unstable fixed points: the ``antichiral'', Gaussian and  O($2N$) Wilson-Fisher fixed points. The chiral and antichiral fixed points merge when $N=N_c$, and for $N<N_c$ there is no stable fixed point so that the transition is first order. There is however a remnant of the fixed points $C^+$ and $C^-$, namely two unphysical fixed points with complex coordinates~\cite{Zumbach93,Delamotte04}. The NPRG approach in the LPA$'$, with the field semi-expansion discussed in Sec.~\ref{subsec_nprg}, gives $N_c\simeq 5.3$ whereas perturbative RG near four dimensions predicts $N_c\simeq 6.2$ in three dimensions. 

Figure~\ref{fig_N65} shows the flow diagram of the O(6)$\times$O(2) model in the plane ($\tlamb_1,\tlamb_2$), obtained from the numerical solution of the NPRG equations. Several critical trajectories ($\mu=\mu_c$) with various initial solutions, including those corresponding to the coupling constants $\lamb_1$ and $\lamb_2$ of $^{87}$Rb, $^{41}$K and $^7$Li, are shown. All critical trajectories are drawn to the chiral fixed point $C^+$ which therefore controls the second-order phase transition. 

In the O(5)$\times$O(2) model there is no stable fixed point but some trajectories, including those corresponding to $^{87}$Rb, $^{41}$K and $^7$Li, are nevertheless drawn to a small region of parameter space where the flow is very slow, before eventually running away as expected for a first-order transition (Fig.~\ref{fig_N65}, right panel). The region where the flow is very slow is located around the projection of the two unphysical fixed points (with complex coordinates) onto the plane ($\tlamb_1,\tlamb_2$). 

\begin{figure}
\centerline{\includegraphics[width=6cm]{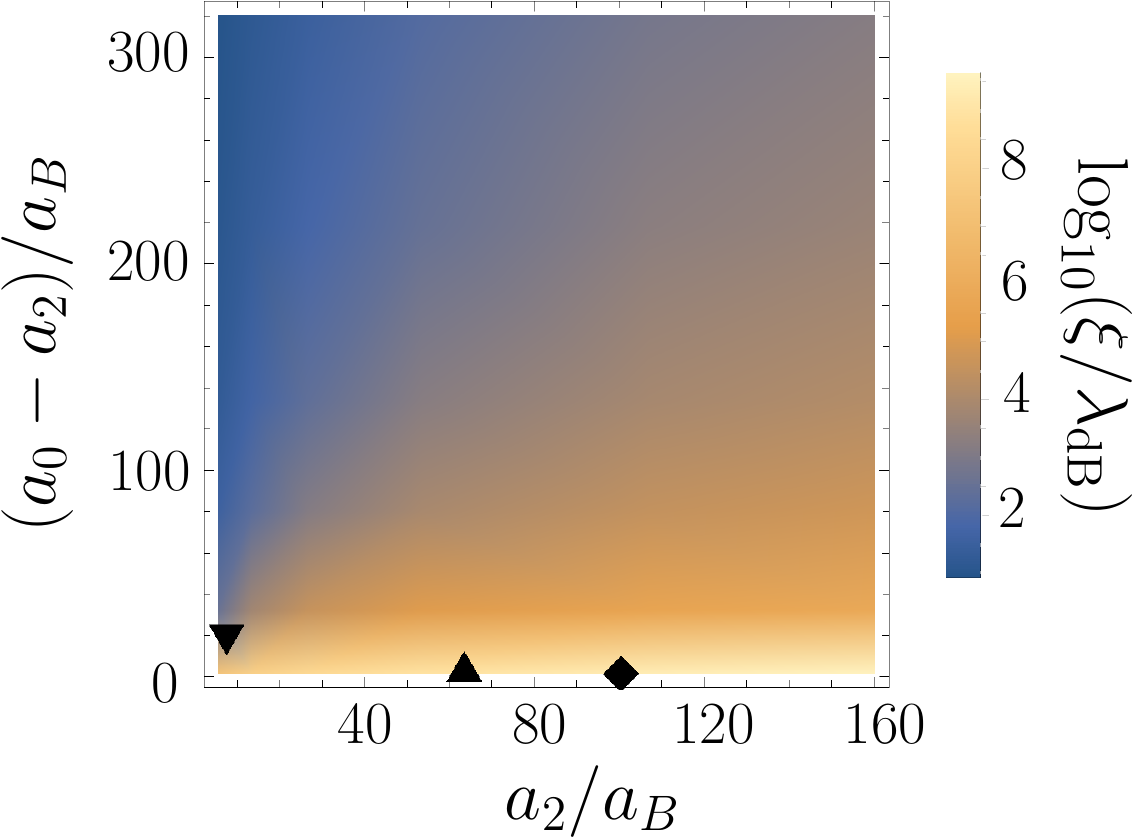}}
\caption{Correlation length of the O(3)$\times$O(2) model at the transition as a function of the scattering lengths $a_0$ and $a_2$ ($\blacklozenge$: $^{87}$Rb, $\blacktriangle$: $^{41}$K, $\blacktriangledown$: $^7$Li).}
\label{fig_xivsa0a2} 
\end{figure}

In the O(4)$\times$O(2) and O(3)$\times$O(2) models (Fig.~\ref{fig_N34}), there is no slowdown of the flow {\it stricto sensu}. However, for trajectories passing by the projection of the unphysical fixed points, the flow remains slow and yields large correlation lengths and pseudoscaling. Not all trajectories pass near the unphysical fixed points and for some trajectories the flow is fast. Furthermore we note that the correlation length at the transition is highly sensitive to the initial conditions as shown in Fig.~\ref{fig_xivsa0a2}. 

\section{O(2)$\times$O(2) model on a lattice}
\label{sec_O2lat} 

Most of the recent numerical studies of STHAs and O($N$)$\times$O(2) models agree on the first-order character of the phase transition. The only exception seems to be the Monte Carlo simulations of the O(2)$\times$O(2) lattice model reported in Ref.~\cite{Calabrese04a} where  the transition was argued to be first order in some parameter range. In this section, we study this model and show that our results are compatible with those of~\cite{Calabrese04a} even though we find the transition to be first order. 

The lattice model studied in Ref.~\cite{Calabrese04a} is defined by the Hamiltonian
\begin{align}
H =& - \beta \sum_{\r,\mubf} (\varphibf_\r \cdot \varphibf_{\r+\mubf} + \psibf_\r \cdot \psibf_{\r+\mubf} ) 
+ \sum_\r \lbrace \varphibf^2_\r + \psibf^2_\r  \nonumber \\ & + A_4 [( \varphibf^2_\r-1)^2 + ( \psibf^2_\r-1)^2 ] + 2 A_{22} \varphibf^2_\r \psibf^2_\r \rbrace ,
\label{hamO2lat} 
\end{align}
where $\varphibf$ and $\psibf$ are two-component real fields, $\{\r\}$ denotes the $L^3$ sites of a cubic lattice, and $\mubf=\hat\x,\hat\y,\hat\z$ is a unit vector. The model~(\ref{hamO2lat}) is a lattice discretization of the O(2)$\times$O(2) Hamiltonian. On the basis of Monte Carlo simulations with up to $120^3$ lattice sites, the authors of Ref.~\cite{Calabrese04a} argued that the transition is first order for $A_{22}>A^*_{22}=1.52(6)$ (with $A_4=1$) but becomes second order when $A_{22}<A^*_{22}$. (Note that for $A_{22}=A_4$, the symmetry is enlarged to O(4); the transition is second order and controlled by the O(4) Wilson-Fisher fixed point.) 

In the continuum limit, the Hamiltonian~(\ref{hamO2lat}) is equivalent to the O($N$)$\times$O(2) model (with $N=2$) defined in the previous sections if we identify~\footnote{To obtain Eqs.~(\ref{Hmapping}) we use the equivalence between Eqs.~(3.6-3.7) and (4.1) of Ref.~\cite{Calabrese04a}.}  
\begin{equation}
\begin{gathered} 
r = -6 + \frac{2-4A_4}{\beta} , \\ 
\lamb_1 = \frac{8}{\beta^2} A_4 , \quad  
\lamb_2 = \frac{4}{\beta^2} (A_{22}-A_4 ). 
\end{gathered}
\label{Hmapping}
\end{equation}
To mimic the presence of the lattice we introduce an upper momentum cutoff $\qmax$. Requiring the number of degrees of freedom to be conserved would give $\frac{4}{3}\pi\qmax^3=(2\pi)^3$, where $(2\pi)^3$ is the volume of the first Brillouin zone of the cubic lattice (setting the lattice spacing to unity). In practice, we fix the value of $\qmax$ by requiring the critical value of $\beta$ at the transition to be the same in the Monte Carlo simulations of the lattice model and the NPRG analysis of the O(2)$\times$O(2) model in the continuum limit. In the following, we restrict ourselves to the case $A_4=1$ and $A_{22}=7/5<A^*_{22}$, for which the transition was argued to be second order~\cite{Calabrese04a}. Choosing $\qmax\simeq 1.23(6\pi^2)^{1/3}$ we reproduce the value $\beta_c\simeq0.76615$ obtained in~\cite{Calabrese04a}. In the NPRG analysis, we take a large initial momentum cutoff $\Lamb_T=100\gg \qmax$ in order to ensure that mean-field theory is nearly exact for $k=\Lamb_T$~\cite{Dupuis08,Machado10}.   

\begin{figure}
\centerline{\includegraphics[width=6.3cm]{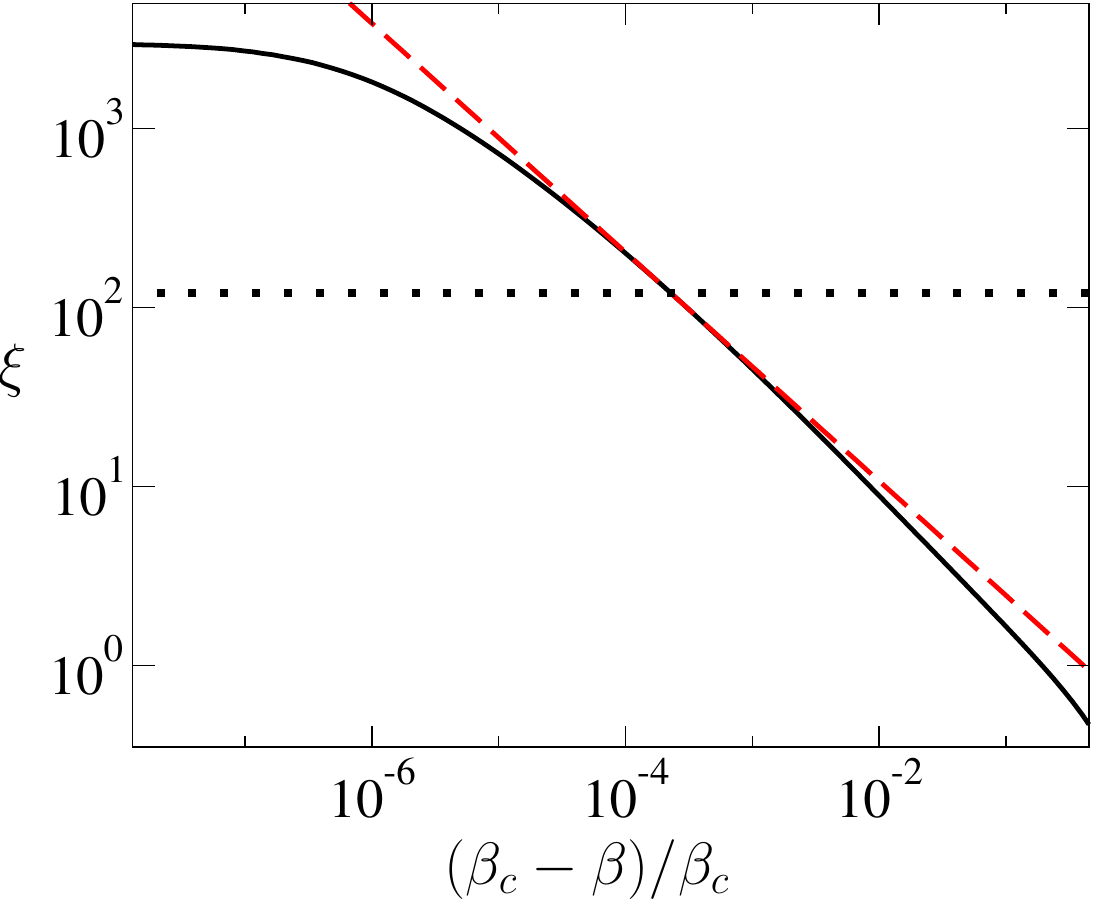}}
\caption{Correlation length $\xi$ (in unit of the lattice spacing) obtained from the NPRG analysis of the O(2)$\times$O(2) lattice model~(\ref{hamO2lat}) when $A_4=1$ and $A_{22}=7/5$. The (red) dashed line corresponds to the fit $\xi\propto (\beta_c-\beta)^{-\nu}$ with $\nu=0.64$. The horizontal dotted line corresponds to a length $L=120$.} 
\label{fig_xiN2} 
\centerline{\includegraphics[width=6.3cm]{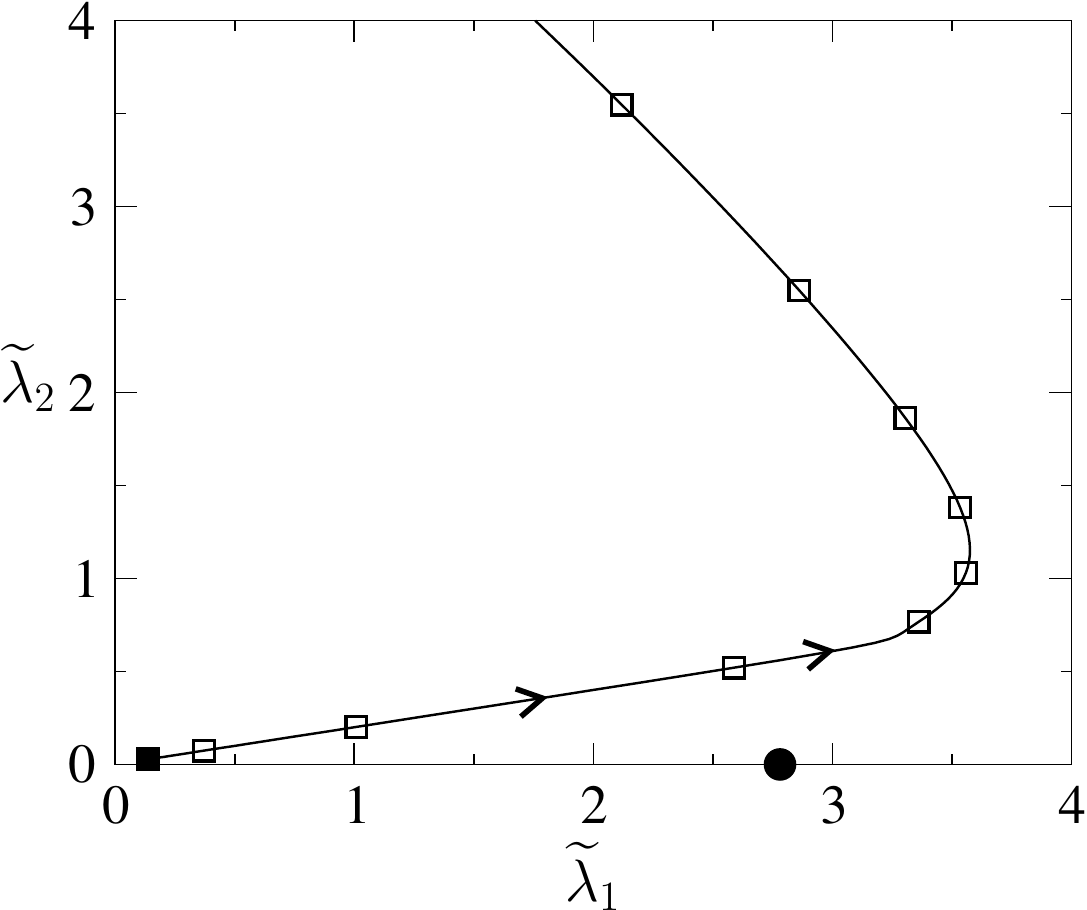}}
\caption{Critical RG flow diagram in the plane $(\tlamb_{1},\tlamb_{2})$ of the O(2)$\times$O(2) lattice model. The solid symbol shows the initial conditions of the trajectory defined by $A_4=1$ and $A_{22}=7/5$. The black dot shows the Wilson-Fisher fixed point of the three-dimensional O(4) model.}
\label{fig_flowO2} 
\end{figure}

Figure~\ref{fig_xiN2} shows the correlation length $\xi$ of the O(2)$\times$O(2) lattice model~(\ref{hamO2lat}) obtained from the NPRG analysis. $\xi$ is large at the transition but finite: the transition is weakly first order. The pseudocritical exponent, defined for a system of size $L=120$~\cite{Calabrese04a}, takes the value $\nu\simeq 0.64$. The value of the pseudocritical exponent $\eta$ (anomalous dimension) is more difficult to estimate since the running anomalous dimension $\eta_k=-k\dk\ln Z_k$ does not reach a quasi-plateau in the flow. If we simply estimate $\eta$ from $\eta_{k=1/L}$ with $L=120$, we find $\eta\simeq0.065$. Thus our results are in reasonable agreement with those
deduced from Monte Carlo simulations, i.e., $\nu=0.63(7)$ and $\eta=0.045(10)$~\cite{Calabrese04a}. The value of $\nu$ is significantly different from the known result $\nu\simeq 0.77$ of the O(4) universality class, in agreement with the fact that the RG trajectory does not pass in the immediate vicinity of the O(4) Wilson-Fisher fixed point (Fig.~\ref{fig_flowO2}). 

We conclude that there is no disagreement between the Monte Carlo simulations and the NPRG analysis of the O(2)$\times$O(2) lattice model (for $A_4=1$ and $A_{22}=7/5$) as far as the value of the (pseudo)critical exponents is concerned. However, in our opinion, the second-order nature of the transition deduced from Monte Carlo simulations is questionable: the weakly first-order nature of the transition, characterized by a large value of the correlation length $\xi\simeq 3\times 10^3$, cannot be inferred from numerical simulations of systems with size $L\ll\xi$ and it is not a surprise that the transition may appear to be second order.

\section{Experimental discussion and conclusion}   

We have shown that phase transitions in spin-one Bose gases and STHAs are described by the same Landau-Ginzburg-Wilson Hamiltonian with O(3)$\times$O(2) symmetry. Spin-one Bose gases, where the low-energy Hamiltonian is fully determined by the boson mass and the scattering lengths $a_0$ and $a_2$, provide us with a test-bed system enabling to test the NPRG predictions regarding the first-order character of the transition as well as pseudoscaling without universality. This opens up the possibility to simulate STHAs with spinor Bose gases and solve the long-standing controversy about the nature (second or weakly first order) of phase transitions in these frustrated magnets~\cite{Debelhoir16a}.

The value of the pseudocritical exponent $\nu$ in $^{87}$Rb and $^{41}$K atom gases, which is close to $\nu_{\rm O(6)}$, is largely a consequence of a crossover phenomenon due to the proximity of the O(6) Wilson-Fisher fixed point, and is independent of the ultimate first-order character of the transition. By contrast, the value $\nu\simeq 0.60$ in $^7$Li is not related in any way to the existence of a nearby critical fixed point but is due to two unphysical fixed points (with complex coordinates) which are a remnant of the chiral and antichiral fixed points in the O($N$)$\times$O(2) model when $N>N_c$ ($N_c\simeq 5.3$). 

We have also considered a O(2)$\times$O(2) lattice model, mimicking the lattice by introducing in our continuum model an upper momentum cutoff of the order of the inverse lattice spacing. We find a weakly first-order transition with values of (pseudo)critical exponents in agreement with lattice Monte Carlo simulations~\cite{Calabrese04a}, although the transition was (wrongly, in our opinion) argued to be second order in this latter study. Our results highlight the difficulty to predict the order of a weakly first-order transition from numerical simulations of systems with size much smaller than the correlation length at the transition.

Our predictions for spin-one Bose gases can be tested by determining experimentally the correlation length $\xi$ and the pseudocritical exponent $\nu$ using matter-wave interferometry~\cite{Donner07,Navon15}. Recent experiments, where the atoms were trapped in a quasi-uniform potential, offer interesting prospects for the measurement of critical exponents in cold atomic gases~\cite{Gaunt13}.
One could also consider varying the scattering lengths $a_0$ and $a_2$ by means of a Feshbach resonance. But the external magnetic field, which in general is used to tune the resonance, would unfortunately suppress the O(3) spin-rotation symmetry. A way out of this difficulty could come from microwave-induced Feshbach resonances as proposed in Ref.~\cite{Papoular10}. Modifying the scattering lengths in $^7$Li would allow a direct confirmation of pseudoscaling, i.e., that the value of $\nu$ changes with $a_0$ and $a_2$. This would also enable us to distinguish our results from the predictions of perturbative RG in fixed dimension $d=3$~\cite{Pelissetto01a,Calabrese02,Calabrese03,Calabrese04a} and conformal bootstrap program~\cite{Nakayama15}, namely a second-order phase transition with a critical exponent $\nu\simeq 0.63$ which turns out to be very close to the value of the pseudocritical exponent $\nu$ in $^7$Li as predicted by NPRG.
Note that increasing $a_0$ by a factor of 4 (with $a_2$ fixed) would be sufficient to make $\xi(\mu_c)$ smaller than the size of the system and thus make the first-order character of the transition observable.

\begin{acknowledgments}
We would like to thank B. Delamotte, D. Mouhanna and M. Tissier for numerous discussions on the NPRG approach to frustrated magnets, and F. Gerbier and C. Salomon for enlightening discussions on spin-one Bose gases.  
\end{acknowledgments}

\appendix

\section{Flow equations in the quantum limit} 
\label{app_rgeq_quantum}

In the quantum limit $k\geq\Lamb_T$, the RG equations of the coupling constants defined in~(\ref{Uvac}) are given by 
\begin{equation}
\begin{aligned}
 \partial_t u_{0,k} ={}& \frac{12 v_d}{d} \frac{k^{2+d}}{2 M} \coth \left(\frac{\eps_k}{2 T}\right), \\
 \partial_t u_{1,k} ={}& -\frac{8 v_d}{d} \frac{k^{2+d}}{2 M} \frac{{u_{2,k}}+{v_{1,k}}}{T} \text{csch}^2\left(
 \frac{\eps_k}{2 T}\right), \\
 \partial_t u_{2,k} ={}& \frac{2 v_d}{d} \frac{k^{2+d}}{2 M} \text{csch}^2\left(\frac{\eps_k}{2 T}\right) \\ & 
 \times \biggl\{ \frac{2}{T^2} \left(3 u_{2,k}^2+4 {u_{2,k}} {v_{1,k}}+8 v_{1,k}^2\right) \coth \left(\frac{\eps_k}{2 T}\right) \\ &  + \frac{u_{2,k}^2}{T \eps_k} \left[ 1 + \frac{T}{\eps_k} \sinh \left(\frac{\eps_k}{T}\right) \right] \biggr\}, \\
   \partial_t v_{1,k} ={}& \frac{4 v_d}{d} \frac{k^{2+d}}{2 M} \frac{{v_{1,k}}}{T} \text{csch}^2\left(\frac{\eps_k}{2 T}\right) \left\{ \frac{2 {u_{2,k}}}{T} \coth \left(\frac{\eps_k}{2 T}\right) \right. \\
   & \left. +\frac{{u_{2,k}}+3 {v_{1,k}}}{\eps_k} \left[ 1 + \frac{T}{\eps_k} \sinh \left(\frac{\eps_k}{T}\right) \right] \right\}.
 \end{aligned}
\label{rgeqQ} 
\end{equation}
where $\eps_k=k^2/2M+u_{1,k}$ and $t=\ln(k/\Lamb)$. For $T=0$, this yields 
\begin{equation}
\begin{aligned}
 \partial_t u_{0,k} ={}& \frac{12 v_d}{d} \frac{k^{2+d}}{2 M}, \\
 \partial_t u_{1,k} ={}& 0, \\
 \partial_t u_{2,k} ={}& \frac{4 v_d}{d} \frac{k^{2+d}}{2 M} \frac{u_{2,k}^2}{\eps_k^2}, \\
   \partial_t v_{1,k} ={}& \frac{8 v_d}{d} \frac{k^{2+d}}{2 M} \frac{v_{1,k} (u_{2,k}+3 v_{1,k})}{\eps_k^2}.
 \end{aligned}
\label{rgeqQzeroT} 
\end{equation}
and we recover, when $u_{1,k}=0$, the expressions of $g_{2,k}=u_{2,k}$ and $g_{0,k}=6v_{1,k}-u_{2,k}$ given in~(\ref{gFkvac}).

\subsection*{Initial conditions in the O(3)$\times$O(2) model} 

The solution of RG equations~(\ref{rgeqQ}) at scale $k=\Lamb_T$ yields the initial conditions of the classical spin-one-boson Hamiltonian~(\ref{ham4}) or, equivalently, the O(3)$\times$O(2) model Hamiltonian~(\ref{ham2}). In practice, it is convenient to express all lengths in unit of the momentum cutoff $\Lamb_T$. One then obtains Hamiltonian~(\ref{ham2}) with 
\begin{equation}
\begin{split} 
r &= \frac{2M}{\Lamb_T^2} u_{1,\Lamb_T} , \\ 
\lamb_1 &= \frac{4M^2}{\beta\Lamb_T} g_{2,\Lamb_T} , \\
\lamb_2 &= \frac{4M^2}{3\beta\Lamb_T} (g_{0,\Lamb_T}-g_{2,\Lamb_T}) , 
\end{split}
\end{equation}
and a unit cutoff.

\subsubsection*{Condensate-density jump}

At the transition the jump $\Delta n_0$ of the condensate density is then related to the jump $\Delta\rho_0$ of the order parameter of the O(3)$\times$O(2) model~(\ref{ham2}) by 
\begin{equation}
\Delta n_0 = \frac{4\pi\Lamb_T}{\lamb^2_{\rm dB}} \Delta\rho_0. 
\end{equation} 
Numerically, one finds that $\xi\Delta\rho_0$ is a number of order unity, with a universal limit ($\simeq 0.4$) when $\lamb_2/\lamb_1\to 0$. Here $\xi$ denotes the correlation length in the model~(\ref{ham2}) with all lengths measured in units of $\Lamb^{-1}_T$. A similar result was obtained in the U($N$)$\times$U($N$) model~\cite{Berges97a}. We deduce 
\begin{equation}
\lamb^3_{\rm dB} \Delta n_0 \sim \frac{\lamb_{\rm dB}}{\xi} , 
\end{equation}
where $\xi$ is now the dimensionful correlation length of the bosonic model~(\ref{ham4}).
Since $n\lamb^3_{\rm dB}\sim 1$ at the transition, one finally obtains
\begin{equation}
\frac{\Delta n_0}{n} \sim \frac{\lamb_{\rm dB}}{\xi} . 
\label{dn0}
\end{equation}
The proportionality coefficient in~(\ref{dn0}) becomes universal ($\simeq 0.6$) when $\lamb_2/\lamb_1\to 0$.

\section{Flow equations in the O($N$)$\times$O(2) model} 
\label{app_rgeq_ONO2}

To derive the flow equations of the classical O($N$)$\times$O(2) model, we consider the following uniform field configuration,
\begin{equation}
 \phibf_1=
 \begin{pmatrix}
 \phi_\alpha \\
 0 \\
 0 \\
 \vdots
\end{pmatrix}
\quad \text{and} \quad \phibf_2=
 \begin{pmatrix}
 0 \\
 \phi_\beta \\
 0 \\
 \vdots
\end{pmatrix} ,
\label{rgeq2}
\end{equation}
where the 2-point vertex takes the form 
\begin{equation}
\Gamma^{(2)}_{a_1,i_1;a_2,i_2}({\bf p},-{\bf p})=
\begin{pmatrix}
 A & 0 & 0 & C &   &   &   &   & \\
 0 & E & D & 0 &   &   &   &   & \\
 0 & D & F & 0 &   &   & \mbox{\Large 0} &   & \\
 C & 0 & 0 & B &   &   &   &   & \\
   &   &   &   & G &   &   &   & \\
   &   &   &   &   & H &   &   & \\
   &   & \mbox{\Large 0}  &   &   &   & G &   & \\
   &   &   &   &   &   &   & H & \\
   &   &   &   &   &   &   &   & \ddots
\end{pmatrix} ,
\label{rgeq1}
\end{equation}
with $a_j=1\cdots N$ and $i_j=1,2$. The first line in~(\ref{rgeq1}) corresponds to the $(1,1;1,1),(1,1;1,2),(1,1;2,1)\cdots (1,1;N,2)$ matrix elements and we have defined 
\begin{align}
A={}& Z \p^2+U^{(0)}{}'+2 \sqrt{\tau} {U^{(1)}} \nonumber \\ & + \phi_\alpha^2 \bigl( U^{(0)}{}'' + 4 \sqrt{\tau} U^{(1)}{}' + 4 \tau U^{(2)} + 2 U^{(1)} \bigr) , \nonumber \\
B={}& Z \p^2+U^{(0)}{}'-2 \sqrt{\tau} U^{(1)} \nonumber \\ & + \phi_\beta^2 \bigl( U^{(0)}{}'' - 4 \sqrt{\tau} U^{(1)}{}' + 4 \tau U^{(2)} + 2 U^{(1)} \bigr) , \nonumber \\
C={}& \phi_\alpha \phi_\beta \bigl( U^{(0)}{}'' - 4 \tau U^{(2)} - 2 U^{(1)} \bigr) , \nonumber \\
D={}& 2 \phi_\alpha \phi_\beta U^{(1)} , \\
E={}& Z \p^2+U^{(0)}{}'-2 \sqrt{\tau} U^{(1)} + 2 \phi_\alpha^2 U^{(1)} , \nonumber \\
F={}& Z \p^2+U^{(0)}{}'+2 \sqrt{\tau} U^{(1)} + 2 \phi_\beta^2 U^{(1)} , \nonumber \\
G={}& Z \p^2+U^{(0)}{}'+2 \sqrt{\tau} U^{(1)} , \nonumber \\
H={}& Z \p^2+U^{(0)}{}'-2 \sqrt{\tau} U^{(1)}, \nonumber 
\end{align}
where $\rho=\half(\phi^2_\alpha+\phi_\beta^2)$ and $\tau=\quarter(\phi^2_\alpha-\phi_\beta^2)^2$. To alleviate the notations, we do not write the dependence on $k$. 

The propagator is diagonal for $\phi_\alpha=\phi_\beta=0$ with all diagonal elements equal to $Z\p^2+U^{(0)}{}'$, which shows that the correlation length in the normal phase is given by~(\ref{xidef}). 

The flow equations of the effective potential and the field renormalization factor are then obtained from 
\begin{equation}
\begin{gathered} 
\dt U(\rho,\tau) = \frac{1}{V} \dt \Gamma[\phibf]\Bigl|_{\phibf\;\rm unif.} \; , \\ 
\dt Z = \lim_{\p\to 0} \frac{\partial}{\partial \p^2} \Gamma^{(2)}_{3,1;3,1}[\p,-\p;\phibf] \Bigl|_{\phibf\;\rm unif.} \; , 
\end{gathered} 
\end{equation}
where the uniform field configuration is defined by~(\ref{rgeq2}), and $\dt Z$ should be evaluated at the minimum of the effective potential, i.e., for $\rho=\rho_0$ and $\tau=0$. This leads to
\begin{widetext} 
\begin{equation}
\begin{aligned}
\partial_t U ={}& \frac{4 v_d}{d (2+d)} Z k^{2+d} (2+d-\eta ) \Bigl\{ (N-2) \frac{2}{D_1} \left(Z k^2+U^{(0)}{}' \right) 
   +\frac{2}{D_2} \left(Z k^2+U^{(0)}{}'+2 \rho U^{(1)} \right) \\
   &+\frac{2}{D_3} \left[ Z k^2+U^{(0)}{}'+2 \rho U^{(1)}+\rho U^{(0)}{}''+4 \tau  \left(U^{(1)}{}'+\rho  U^{(2)}\right) \right] \Bigr\},
\end{aligned}
\label{Uflow} 
\end{equation}
and 
\begin{equation}
\begin{aligned}
\partial_t Z ={}& -\frac{16 v_d}{d} Z^2 k^{2+d} {\rho} \frac{1}{D_4} \Big[ {U^{(0)}{}''}^2 \left(Z k^2+U^{(0)}{}'\right) \left(Z k^2+U^{(0)}{}'+8 {\rho} U^{(1)}\right) \\
   &+ 8 {U^{(1)}}^2 \left(Z k^2+U^{(0)}{}'+2 {\rho} U^{(0)}{}''\right)^2+16 \rho^2 {U^{(0)}{}''}^2 {U^{(1)}}^2 \Bigr]\Bigr|_{\rho=\rho_0,\tau=0} 
\end{aligned}
\end{equation}
where $v_d^{-1}=2^{d+1}\pi^{d/2}\Gamma(d/2)$, $d=3$ is the space dimension, and  
\begin{equation}
\begin{aligned}
D_1={}& \left(Z k^2+U^{(0)}{}'\right)^2-4 \tau  {U^{(1)}}^2 \\
D_2={}& \left(Z k^2+U^{(0)}{}'\right) \left(Z k^2+U^{(0)}{}'+4 \rho  U^{(1)}\right)+4 \tau {U^{(1)}}^2 \\
D_3={}& \left(Z k^2+U^{(0)}{}'+4 \rho  U^{(1)}\right) \left(Z k^2+U^{(0)}{}'+2 \rho U^{(0)}{}''\right) 
     +4 \tau  \left[ 2 Z k^2 \left(U^{(1)}{}'+\rho U^{(2)} \right) +2 U^{(0)}{}' \left( U^{(1)}{}'+\rho  U^{(2)} \right) \right. \\
    &\left. -3 U^{(1)} \left( U^{(0)}{}''+U^{(1)} \right) -4 \rho U^{(1)}{}' \left( U^{(1)}+\rho U^{(1)}{}' \right) +4 \rho ^2 U^{(0)}{}'' U^{(2)}\right] 
    -16 \tau ^2 \left[U^{(2)} \left( U^{(1)}+U^{(0)}{}'' \right)-{U^{(1)}{}'}^2\right]\\
D_4={}& D_1 D_3^2  .
\end{aligned}
\end{equation}
\end{widetext}
Flow equations for $U^{(0)}$, $U^{(1)}$ and $U^{(2)}$ can be deduced from~(\ref{Uflow}) using~(\ref{Uexpand}).  

For the numerical solution of the flow equations, it is convenient to introduce a dimensionless effective potential defined by $\tilde U(\trho,\tilde\tau)=k^{-3}U(\rho,\tau)$, $\trho=Zk^{2-d}\rho$ and $\tilde\tau=Z^2k^{4-2d}\tau$.  
 
\section{Ferromagnetic transition without BEC} 
\label{sec_ferro} 

In this Appendix we show how Eq.~(\ref{critferro}) is obtained. The BEC temperature of a noninteracting spin-one Bose gas is given by 
\begin{equation}
T_c^0 = \frac{2\pi}{M} \left( \frac{n}{3\zeta(3/2)} \right)^{2/3} .
\label{Tbec} 
\end{equation}
The random-phase approximation of Ref.~\cite{Natu11} predicts a ferromagnetic transition (without BEC) at a temperature $T_f$ defined by 
\begin{equation}
\frac{T_f-T_c^0}{T_c^0} = -\alpha \left( \third+\frac{a_2-a_0}{a_0+2a_2} \right) (a_0+2a_2) n^{1/3} 
\label{ferrocrit0}
\end{equation}
in the dilute limit $a_Fn^{1/3}\ll 1$, where 
\begin{equation}
\alpha = \frac{8\pi}{[3\zeta(3/2)]^{4/3}} \simeq 1.61 . 
\end{equation}
Note that Eq.~(\ref{ferrocrit0}) differs from the result of~\cite{Natu11}: the coefficient $\alpha$ is smaller by a factor 1/3 and $a_0+2a_2$ appears instead of $a_0$. Together with the shift $\Delta T_c$ of the BEC transition temperature given in Sec.~\ref{subsec_ferrowosf}, Eq.~(\ref{ferrocrit0}) leads to~(\ref{critferro}). 



\begin{thebibliography}{71}
\bibitem [{not({\natexlab{a}})}]{note19}%
  \BibitemOpen
  \href@noop {} {}  \bibinfo {note} {When the
  O(3)$\times$O(2) model is used to describe the transition in
  STHAs~\cite{Kawamura88}, the mean value of the spin is given by $\mean{\S_i}
  = \mean{\varphibf_1} \cos(\Q\cdot\r_i+\theta) + \mean{\varphibf_2}
  \sin(\Q\cdot\r_i+\theta)$. When $\lamb_2>0$, $\mean{\varphibf_1} \perp
  \mean{\varphibf_2}$ in the low-temperature phase and the spin ordering is
  noncollinear.}\BibitemShut {Stop}%
\bibitem [{not({\natexlab{c}})}]{note21}%
  \BibitemOpen
  \href@noop {} {} \bibinfo {note} {We do not expect
  this instability to disappear in more involved approximation schemes such as
  the Blaizot--M\'endez-Galain--Wschebor
  approach~\cite{Blaizot06,Benitez09,Benitez12}, although this issue would
  deserve a separate study.}\BibitemShut {Stop}%
\end{thebibliography}

%

\end{document}